 \documentclass[final,authoryear,5p,times,twocolumn]{elsarticle}
\usepackage{graphicx}
\usepackage{pstricks}
\usepackage{amssymb}
\journal{PSS}

  \bibpunct{(}{)}{;}{a}{}{,}
\usepackage{hyperref}
\hypersetup{
    unicode=false,          % non-Latin characters in Acrobat's bookmarks
    pdftoolbar=true,        % show Acrobat's toolbar?
    pdfmenubar=true,        % show Acrobat's menu?
    pdffitwindow=false,     % page fit to window when opened
    pdftitle={Shape modeling technique KOALA validated by ESA Rosetta at (21) Lutetia},
    pdfauthor={Benoit Carry},
    pdfsubject={Planetary Science},
    pdfkeywords={},         % list of keywords
    pdfnewwindow=true,      % links in new window
    colorlinks=true,        % false: boxed links; true: colored links
    linkcolor=gray,         % color of internal links
    citecolor=blue,         % color of links to bibliography
    filecolor=gray,         % color of file links
    urlcolor=gray           % color of external links
}

\usepackage{ulem}

\newcommand{\add}[1]{\textbf{#1}}
\newcommand{\rem}[1]{\textcolor{red}{\sout{#1}}}

\renewcommand{\add}[1]{#1}
\renewcommand{\rem}[1]{}

\newcommand{\arcsec}{\ensuremath{^{\prime\prime}}}
\newcommand{\degr}{\ensuremath{^\circ}}
\newcommand{\ie}{\textsl{i.e.}}
\newcommand{\eg}{\textsl{e.g.}}

\begin{document}

\begin{frontmatter}



\title{Shape modeling technique KOALA validated by ESA Rosetta at (21) Lutetia}


\author[esac]{B. Carry}
\ead{benoit.carry@esa.int}
\author[tut]{M. Kaasalainen}
\author[swri]{W. J. Merline}
\author[mpe]{T. G. M\"uller}
\author[lam]{L. Jorda}
\author[aflr]{J. D. Drummond}
\author[imcce]{J. Berthier}
\author[esac]{L. O'Rourke}
\author[praha]{J. \v{D}urech}
\author[esac]{M. K{\"u}ppers}
\author[mpi]{A. Conrad}
\author[swri]{\add{P. Tamblyn}}
\author[eso]{C. Dumas}
\author[mps]{H. Sierks}
\author{and the OSIRIS Team\fnref{osiris}}

\address[esac]{European Space Astronomy Centre, ESA, P.O. Box 78, 28691 Villanueva de la Ca\~{n}ada, Madrid, Spain}
\address[tut]{Tampere University of Technology, P.O. Box 553, 33101 Tampere, Finland}
\address[swri]{Southwest Research Institute, 1050 Walnut St. \#300,
  Boulder, CO 80302, USA}
\address[mpe]{Max-Planck-Institut f\"ur extraterrestrische Physik (MPE), Giessenbachstrasse, 85748 Garching, Germany}
\address[lam]{Laboratoire d'Astrophysique de Marseille, Universit\'e de Provence, Marseille, France}
\address[aflr]{Starfire Optical Range, Directed Energy Directorate, Air Force
    Research Laboratory, Kirtland AFB, NM 87117-577, USA}
\address[imcce]{Institut de M\'ecanique C\'eleste et de Calcul des \'Eph\'em\'erides, Observatoire de Paris, UMR8028 CNRS, 77 av. Denfert-Rochereau 75014 Paris, France}
\address[praha]{Astronomical Institute, Faculty of Mathematics and
  Physics, Charles University in Prague, V Hole\v{s}ovi\v{c}k\'ach 2,
  18000 Prague, Czech Republic}
\address[mpi]{Max Planck Institute f\"ur Astronomy (MPA), K\"{o}nigstuhl 17, 69117 Heidelberg, Germany}
\address[eso]{European Southern Observatory, Alonso de C\'{o}rdova 3107, Vitacura, Casilla 19001, Santiago de Chile, Chile}
\address[mps]{Max-Planck-Institut f\"ur Sonnensystemforschung (MPS), Max-Planck-Strasse 2, 37191 Katlenburg-Lindau, Germany}

\fntext[osiris]{%
  M. A'Hearn,
  F. Angrilli,
  C. Barbieri, 
  A. Barucci,
  J.-L. Bertaux,
  G. Cremonese,
  V. Da Deppo,
  B. Davidsson,
  S. Debei,
  M. De Cecco,
  S. Fornasier,
  M. Fulle, 
  O. Groussin, 
  P. Guti\'errez,
  W.-H. Ip,
  S. Hviid,
  H. U. Keller,
  D. Koschny, 
  J. Knollenberg,
  J. R Kramm,
  E. Kuehrt,
  P. Lamy, 
  L. M. Lara,
  M. Lazzarin,
  J. J. L\'opez-Moreno,
  F. Marzari,
  H. Michalik,
  G. Naletto,
  H. Rickman, 
  R. Rodrigo, 
  L. Sabau,
  N. Thomas,
  K.-P. Wenzel}

\begin{abstract}
  We present here a comparison of our results from
  ground-based observations of asteroid (21) Lutetia with
  imaging data acquired during the flyby of the asteroid by
  the ESA Rosetta mission.
  This flyby provided a unique opportunity to evaluate and 
  calibrate our method of determination of size, 3-D shape, and spin of
  an asteroid from ground-based observations.
  Knowledge of certain observable physical properties of small bodies
  (\eg, size, spin, 3-D shape, and density) have
  far-reaching implications in furthering our
  understanding of these objects, such as
  composition, internal structure, and the effects of
  non-gravitational forces.
  We review the different observing techniques used to
  determine the above physical properties of asteroids
  and present our 3-D shape-modeling technique KOALA
  -- Knitted Occultation, Adaptive-optics, and Lightcurve Analysis --
  which is based on multi-dataset inversion.
  We compare the results we obtained with KOALA, prior to the flyby, on asteroid
  (21) Lutetia with the high-spatial resolution images of the asteroid taken
  with the
  OSIRIS camera on-board the ESA Rosetta spacecraft, during its encounter
  with Lutetia on 2010 July 10.
  The spin axis determined with KOALA was found to be accurate
  to within two degrees, while the KOALA
  diameter determinations were within 2\% of the Rosetta-derived
  values. 
  The 3-D shape of the KOALA model is also confirmed by the
  spectacular visual agreement between both
  3-D shape models (KOALA pre- and OSIRIS post-flyby).
  We found a typical deviation of only 2\,km at local scales between
  the profiles from KOALA predictions and OSIRIS images, resulting
  in a volume uncertainty provided by KOALA \add{better than 10\%}.
%  in an upper limit of about 15\% for the volume uncertainty provided by KOALA.
%
  Radiometric techniques for the interpretation of thermal infrared
  data also benefit greatly 
  from the KOALA shape model: the absolute
  size and geometric albedo can be derived with high accuracy, and
  thermal properties, for example the thermal inertia,
  can be determined unambiguously.
  The corresponding Lutetia analysis leads
  to a geometric
  albedo of 0.19\,$\pm$\,0.01
  and a thermal inertia below 40\,J\,m$^{-2}$\,s$^{-0.5}$\,K$^{-1}$,
  both in excellent agreement with the Rosetta findings.
  We consider this to be a validation of the KOALA method.
  Because space exploration will remain limited to only a few objects,
  KOALA stands as a powerful technique to study a much larger set
  of small bodies using Earth-based observations.
\end{abstract}


\end{frontmatter}

% \linenumbers
% \pagewiselinenumbers
\section{Remote-sensing shape modeling\label{sec: intro}}
  \indent Perhaps the most striking observable of any asteroid is its shape. 
  In 1993, spacecraft exploration revealed for the first time the
  stunning
  non-spherical shape of asteroid (951) Gaspra when NASA's Galileo spacecraft
  made the first of its two asteroid encounters,
  on its way to Jupiter. Asteroids had remained point-sources in the
  sky since the discovery of (1) Ceres in 1801 by Piazzi, almost two
  centuries before.
  Only the advent of space exploration and large Earth-based
  telescopes
  (\eg, Arecibo and Goldstone radio telescopes,
  space-based-optical -- HST,
  or ground-based near-IR, equipped with adaptive optics
  -- \add{Palomar,} Lick, CFHT, Keck, ESO VLT, Gemini) 
  allowed their apparent disks to be spatially resolved, and
  their irregular shapes to be imaged.
  The past decade has seen a revolution in our understanding of the
  physical properties (\eg, size, 3-D shape, spin axis) of asteroids.
  This revolution has come about
  thanks to improved observing facilities and, of equal
  importance, from improved methods of analysis. \\ 
%%%
  \indent Determination of the physical properties for a
  statistically relevant set
  of asteroids has many implications for our
  understanding of these remnants of solar-system formation and, in turn,
  can be expected to improve our understanding of the history and evolution of
  the Solar System.
  For instance, the distribution of spin axes of the larger asteroids 
  (diameter larger than $\sim$100 km) on the celestial sphere
  is not expected to be isotropic.
  Numerical hydrocode simulations have predicted a
  slight excess in prograde rotators, due to the gas-pebble
  interaction in the protoplanetary disk 
  \citep{2010-MNRAS-404-Johansen}.
  Similarly, the spin state of \textsl{small} asteroids
  (diameter not larger than few kilometers)
  is dominated by the non-gravitational YORP effect
  \citep{2011-AA-530-Hanus}.
  Statistical knowledge of spin coordinates, how they
  are distributed within and among
  asteroid families of different ages,
  will provide strong constraints on the
  effectiveness of YORP \citep{2003-Icarus-162-Slivan}.\\
  \indent Reconstruction of the 3-D shape (including the size) is
  required to estimate the volume of an asteroid,
  which in turn is used to derive its density, possibly
  the property most fundamental 
  to our understanding of
  an asteroid \citep{2002-AsteroidsIII-4.2-Britt}.
  Observations of the surface of an asteroid, such as
  \add{colors}, spectra, or phase effects, can provide clues to the 
  surface composition of the asteroid.  This may or may not
  be related to the bulk composition of the body
  \citep[\eg,][]{2011-EPSL-305-Elkins-Tanton}.
  Masses for asteroids can be determined from a spacecraft flyby,
  from the orbital motion of a natural moon, or even from the
  perturbations of asteroids on other bodies, such as Mars
  \citep{2002-AsteroidsIII-2.2-Hilton}.
  In most cases, however, the uncertainty in the density is dominated by
  the uncertainty in the volume, rather than the uncertainty in the
  mass \citep{2002-AsteroidsIII-2.2-Merline}.
  Precise reconstruction of the 3-D shape is therefore
  of high importance for all asteroids 
  for which a mass has been, or will be,
  estimated \citep[\eg,][]{2002-AsteroidsIII-2.2-Hilton,
    2007-AA-472-Mouret, 2011-AJ-141-Baer}. \\
  \indent From the comparison of an asteroid's density with the
  densities of its 
  most-likely constituents, we can constrain the macroporosity
  \rem{(the presence of, and extent of, large-scale voids)}
  \add{(large-scale voids)}
  in its interior,
  probably produced by impacts over its history
  \citep{2002-AsteroidsIII-4.2-Britt}.
  These impacts could have partially disrupted the body,
  producing large-scale fractures, or even totally disrupted
  the body, with subsequent reaccumulation of the resulting
  fragments, leading to a ``rubble-pile'' structure.\\
  \indent Evidence of gigantic,
  but less than totally disruptive, impacts can 
  be seen by high-resolution imaging and also inferred from our
  shape-modeling of asteroids. The huge impact craters 
  evident in the images of C-type asteroid (253) Mathilde
  \citep{1997-Science-278-Veverka}
  are thought to be about as large as could
  be sustainable by a body without disruption. 
  There have been suggestions
  that these craters were created by compaction of low-density
  target material, rather than explosive ejection typical of
  hard-rock impacts \citep{1999-Nature-402-Housen}.
  Already, our ground-based adaptive-optics imaging has shown
  what appear to be facets or depression-like features, similar to
  those seen on Mathilde, in
  some other large C-type asteroids
  \citep[\eg, (511) Davida in][]{2007-Icarus-191-Conrad}.
  Alternatively, some of our other images of C-type asteroids, such as
  (52) Europa, appear to bear no evidence of giant impacts
  \citep[see][]{2011-Icarus--Merline}.
  Evaluation of the prevalence
  of such large impact events can give us insight into the size and
  frequency of these impact events over time, and thus into the
  history of the impacting population.\\
  \indent We summarize below the most common of the many
  observing techniques used to derive size, 3-D shape, and spin-vector
  coordinates
  and highlight some of their advantages and drawbacks.
  Then, in Section~\ref{sec: koala}, we describe our KOALA multi-data
  shape-modeling algorithm.
  In Section~\ref{sec: comp}, we present a
  comparison of the results produced by
  KOALA (from Earth-based observations) with those derived from the
  ESA Rosetta flyby of asteroid (21) Lutetia.
  In Section~\ref{sec: thermal}, we use our KOALA model 
  \add{in conjunction with mid-infrared data and a thermophysical
    model}
  to derive the 
  thermal properties of Lutetia and compare the results with those
%  \rem{obtained using flyby constraints}
  \add{derived using thermal observations from the Rosetta spacecraft
  \citep{2011-PSS--Gulkis}
  and from the ground, making use of the shape model from the
    flyby \citep{2011-PSS--Rourke}.} 
  We \add{assess} the accuracy of the KOALA shape-modeling
  method in Section~\ref{sec: conclu}.

  \subsection{Optical lightcurve\label{ssec: lc}}
\begin{figure}
  \includegraphics[width=.5\textwidth]{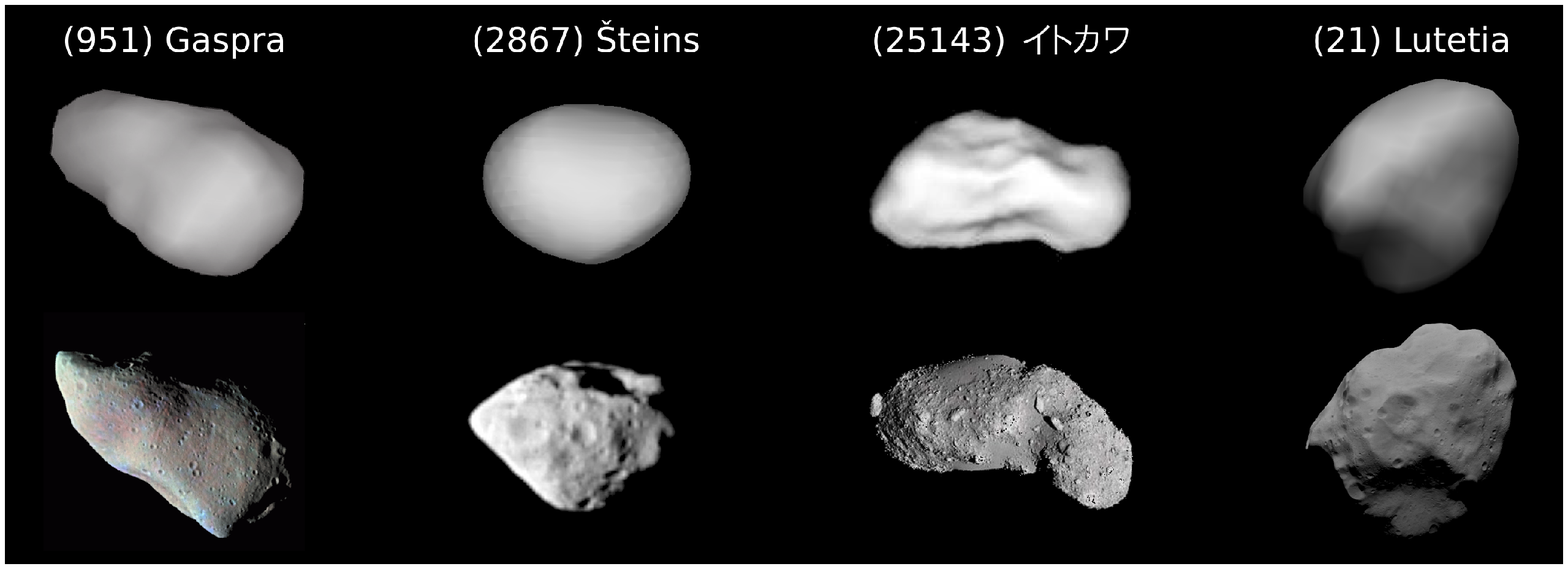}
  \caption{%
    Comparison of four shape models, derived using ground-based
    observations, with high-spatial-resolution
    images obtained \textsl{in situ} by spacecraft (left-to-right):
    (951) Gaspra lightcurve inversion model from
    \citet{2002-AsteroidsIII-2.2-Kaasalainen}, 
    image from NASA Galileo; 
    (2867) \v{S}teins lightcurve inversion model from
    \citet{2008-AA-487-Lamy-b, 2008-AA-487-Lamy-a},
    image from ESA Rosetta;
    (25143) Itokawa radar delay-Doppler model from
    \citet{2005-MPS-40-Ostro}, 
    image from JAXA Hayabusa;
    and
    (21) Lutetia KOALA model from
    \citet{2010-AA-523-Carry} and \citet{2010-AA-523-Drummond},
    image from ESA Rosetta.
    \label{fig: models}
  }
\end{figure}

    \indent Historically, spin properties 
    and triaxial-ellipsoid shapes have been
    studied largely through observations of
    \add{rotationally induced variability in disk-integrated brightness}
    \rem{disk-integrated photometric variations during rotation}
    (lightcurves).
    Indeed, the object's
    shape, its rotational state (period and spin-vector coordinates),
    and the scattering properties
    of its surface can be determined from the analysis of its lightcurves
    over time (as the viewing/illumination geometry changes).\\
    \indent For about a decade, starting with
    the lightcurve inversion algorithm presented by 
    \citet{2001-Icarus-153-Kaasalainen-a} and
    \citet{2001-Icarus-153-Kaasalainen-b},
    lightcurves also have been used extensively
    to derive 3-D shape models of asteroids
    (see the examples for asteroids Gaspra and \v{S}teins in
    Fig.~\ref{fig: models}). 
    The 3-D shape models and spin properties of more than 200
    asteroids already have been derived
    \citep[these are accessible from
      DAMIT\,\footnote{DAMIT:~\href{http://astro.troja.mff.cuni.cz/projects/asteroids3D/web.php}{http://astro.troja.mff.cuni.cz/projects/asteroids3D/web.php}},
      see][]{2010-AA-513-Durech}.
    These shape models are, however, limited to dimensionless,
    \textsl{convex} shapes, with limited spatial resolution.
    Recently, 
    \citet{2011-Icarus-214-Durech} have shown that
    the size of these models can be set by using
    other types of input data
    (\eg, stellar occultation profiles, see Sect.~\ref{ssec: occ}).
    The intrinsic \textsl{convex} nature of the models
    precludes accurate determination of the volume, and hence density, of the
    objects, however. \\ 
    \indent Because lightcurve observations require
    neither large telescope aperture
    nor specialized instrumentation, they
    are, and will remain, a major source of information on
    small bodies.
    Thousands of lightcurves, for hundreds of
    asteroids\,\footnote{\eg,
      the Asteroid Photometry Catalogue (APC)
      or the Asteroid Lightcurve Database \citep[LCDB,][]{2009-Icarus-202-Warner}
      list
      more than 6000 lightcurves for about 700 asteroids: \\
      APC: \href{http://asteroid.astro.helsinki.fi/apc}{http://asteroid.astro.helsinki.fi/apc}\\
      LCDB: \href{http://www.minorplanet.info/lightcurvedatabase.html}{http://www.minorplanet.info/lightcurvedatabase.html}
    },
    have been accumulated during the last half century.
    Amateur astronomers 
    contribute significantly to this ever-growing database
    of lightcurves\,\footnote{\eg, about 2300
      lightcurves for more than 1700 asteroids have been
      acquired by the CdR group:
      \href{http://obswww.unige.ch/~behrend/page_cou.html}{http://obswww.unige.ch/$\sim$behrend/page\_cou.html}}
    \citep[\eg,][]{2006-AA-446-Behrend, 2007-AA-465-Durech}.\\
    \indent Sparse photometry
    (\ie, when the typical separation between
    measurements is larger than the rotation period, as opposed to
    historical lightcurves, which are dense in time) can also be used to
    reconstruct 3-D models \citep[see][]{2004-AA-422-Kaasalainen}.
    \citet{2011-AA-530-Hanus} have used a combination of sparse
    photometry, together with dense lightcurves,
    to derive about 100 new shape models,
    using measurements extracted from large all-sky surveys
    (such as USNO, Catalina, Siding Spring, and Hipparcos).
    \add{Knowledge of the absolute photometry (as opposed to
      relative photometric measurements, as in a dense lightcurve) 
      is, however, required to make these sparse measurements useful
      to 3-D shape modeling
      \citep[see][for a detailed discussion]{2011-AA-530-Hanus}.
    }
    From the upcoming PanSTARRS and Gaia surveys
    we can expect hundreds of thousands of objects to be modeled using
    this method \citep{2005-EMP-97-Durech}.

  \subsection{Stellar occultation\label{ssec: occ}}
    \indent Occultations occur when a solar-system object passes
    between a star and Earth, causing the asteroid's shadow to cross
    some portion of Earth as a track.  Because the star is very far
    relative to the Earth-asteroid separation, the shadow cast by
    the asteroid is effectively parallel. Thus, the width of the
    shadow track (perpendicular to the track)
    gives the maximum
    width of the asteroid in the cross-track
    direction. It is usually not possible to get a high density
    of observers stationed across the track, and thus this dimension may
    not be so well established. But along the track, the size of
    the asteroid, as an along-track chord on the asteroid, corresponding
    to the position of the observer within the shadow,
    is given by the length of time
    of the blink-out event.  With many observers, many along-track
    events can be recorded and the blink-out intervals are converted
    to chord lengths at the asteroid by using the asteroid's known
    speed
    \citep[see][for a review]{1989-AsteroidsII-Millis}.\\
    \indent One advantage of stellar occultations
    is that very small minor planets can be probed
    (provided one accounts properly for diffraction effects).
    Even stellar occultations by small trans-Neptunian objects (TNOs)
    of a few kilometers diameter can be recorded
    \citep[see][]{2009-EMP-105-Roques}.
    Finally, stellar occultations provide a powerful means for studying
    thin atmospheres and/or exospheres
    \citep[\eg,][]{2003-Nature-424-Sicardy}.\\ 
    \indent In theory, three occultation events
    (each recorded by enough observers
    to secure a unique
    apparent-ellipse to be fit on the chords)
    provide enough constraints to
    determine the triaxial dimensions (ellipsoid) and spin-vector coordinates 
    of an asteroid 
    \citep{1989-Icarus-78-Drummond,1990-AJ-99-Dunham}.
    The number of chords that result from an event will often be larger
    with brighter occulted stars. 
    This is because many more
    observers can be fielded if the
    \rem{requirement for telescope aperture}
    \add{required telescope aperture}
    is modest.  In practice, however,
    occultations of bright stars by any
    given asteroid occur rather infrequently. 
    This difficulty in obtaining
    sufficient chords, plus the
    \rem{levels of noise}
    \add{noise level}
    often encountered
    (especially the systematic errors coming from imperfect
    knowledge of asteroid and star astrometry, combined with
    observer timing errors), strongly limit our ability to
    construct
    3-D shape models and derive spin properties from
    occultations alone.
    Nevertheless, stellar occultations are an efficient way to
    provide additional size/shape information, particularly for shape
    models that otherwise lack a scale, such as those from lightcurves alone
    \citep[see~\ref{ssec: lc} and][]{2011-Icarus-214-Durech}.\\
    \indent With the availability of low-cost
    GPS positioning equipment and CCD cameras
    (the majority of measurements are made
    by amateur astronomers),
    the accuracy of occultation timings has improved
    greatly over the last
    decade. From this improved precision, 
    together with the refinement of the orbits of small bodies
    expected to result from Gaia/PanSTARRS
    \citep[allowing an extremely
    precise prediction of the occultation track on Earth,
    see][]{2007-AA-474-Tanga}, 
    we can expect that stellar occultations
    will become ever more useful in the
    determination of certain
    physical properties of asteroids, especially for objects
    having small angular diameters.

  \subsection{Thermal radiometry\label{ssec: thermal}}
    \indent The amount of thermal emission from an asteroid is
    mainly a function of its physical diameter and surface albedo, and,
    to a lesser extent, the physical properties of its surface
    (\eg, thermal inertia, roughness).
    Main-belt asteroids are among the brightest sources
    in the sky in the mid-infrared (5--20\,microns), so infrared
    satellites (IRAS, ISO, AKARI, Spitzer, \add{Herschel,} WISE) have 
    been able to acquire observations of a vast number
    of these objects
    \citep[see][for instance]{2011-ApJ-731-Mainzer}.\\
    \indent Estimates for size and albedo
    are model-dependent, however, and
    several thermal models co-exist,
    from the simple Standard Thermal Model (STM)
    of non-rotating spheres of
    \citet{1986-Icarus-68-Lebofsky}
    to the detailed Thermophysical Model (TPM)
    of \citet{1996-AA-310-Lagerros,1997-AA-325-Lagerros},
    having a complete description of
    3-D shape and surface properties.
    Unfortunately, the systematic uncertainties 
    involved when applying those models
    (resulting from their respective assumptions and approximations)
    are not always properly taken into account in estimating
    error bars, and
    results often differ from one determination to another by more
    than the quoted uncertainties \citep[see Table 3 in][illustrating the
      issue]{2009-PSS-57-Delbo}. 
    For instance, it now seems that the database of
    2\,228\,diameters \citep{2002-AJ-123-Tedesco-a} estimated using the
    overly simple STM with IRAS data
    may be biased by 
    \add{a few percents \citep[see the re-analysis of IRAS data
        by][]{2010-AJ-140-Ryan}.}\\ 
%    \re6--8\% on average \citep[and up to 20\%, see][]{2006-Icarus-185-Marchis}.}\\
%
    \indent The radiometric technique to derive sizes and
    albedos from thermal infrared data benefits greatly from the
    availability of 3-D shape models.
    For example, the absolute size of (25143) Itokawa
    derived by
    \citet{2005-AA-443-Muller}, based on a 3-D shape model
    implementation in a TPM code, agreed \add{to} within 2\% of the
    final in-situ 
    result from the Hayabusa mission
    \citep{2006-Science-312-Fujiwara}.
    In general, mid-infrared observations are highly valuable for
    scaling dimensionless shape models
    (similar to the situation with stellar
    occultations, as mentioned in Sect.~\ref{ssec: occ})
    and mid-infrared data are available for several thousand
    asteroids. These data 
    even allow determination of the most likely
    spin-axis solutions in cases where
    lightcurve inversion techniques lead to more than one possible shape
    and spin-axis solution
    \add{\citep[see, \eg,][for a recent example]{2011-AA-525-Muller}}.
    In cases where the shape-model already comes with size information
    (or alternatively, if many thermal observations are available for
    a given target), it is possible to derive the thermal inertia,
    indicative of the surface characteristics:
    \eg, bare rock, ice, boulders, dust regolith
    \add{\citep[see, for instance][among many
        others]{2005-AA-443-Muller, 2007-Icarus-190-Delbo}}.\\ 
    \indent In some particular cases of extensively observed
    asteroids, the thermal radiometry can also
    provide hints on the 3-D shape, through the measure of the
    apparent projected cross-section of the asteroid on the
    plane of the sky at each epoch.
    By comparing the predicted with observed thermal fluxes of Lutetia
    under many geometries, \citet{2011-PSS--Rourke} have shown that
    adding a hill/plateau, whose size remains within the 
    quoted 3-D shape uncertainty,
    could explain the discrepancies observed for a certain observing
    geometry.

  \subsection{Radar delay-Doppler echoes}
    \indent Radar observations 
    \rem{involve transmitting a signal toward}
    \add{consist in transmitting a radio signal toward}
    the target and recording the echo.
    The distribution of the echo power
    in delay time and Doppler frequency is related to the spin state
    and 3-D shape of the target
    \citep[see the reviews by][]{1989-AsteroidsII-2-Ostro,
      2002-AsteroidsIII-2.2-Ostro}. 
    The time and frequency precision currently achievable (Arecibo,
    Goldstone) allow the study of very small objects, the main limit
    of radar observations being the distance of the target (echo power
    scales inversely with distance to the fourth power).
    This is why most radar studies have concentrated on Near-Earth
    Objects (NEOs), while dedicated observations of Main-Belt
    Asteroids (MBAs) have been more limited
    \citep[see][]{2002-AsteroidsIII-2.2-Ostro}. \\
    \indent The difficulty in constructing 3-D shapes from a
    series of delay-Doppler ``\textsl{images}'' 
    is due to the absence of a direct,
    one-to-one, link between each region of the surface
    with a pixel in delay-Doppler space.
    Indeed, all points 
    situated at the same range from the observer, and moving at the
    same radial velocity (possibly spread over the apparent disk)
    will contribute to a single delay-Doppler pixel. 
    So delay-Doppler images are many-to-one maps of the shape as
    highlighted by \citet{2002-AsteroidsIII-2.2-Ostro}:
    there is no \textsl{a priori} way to determine how many regions
    will contribute to a single pixel, which strongly contrasts with
    the one-to-one mapping (``WYSIWYG'') achieved in
      disk-resolved imaging.\\ 
    \indent Radar echoes remain, however,
    the best way to determine the
    physical properties of NEOs
    (\eg, the very small NEO Itokawa in Fig.\ref{fig: models}).
    For instance, the possible migration of the regolith at the
    surface of a fast-rotating asteroid 
    triggered by YORP spin-up \citep{2008-Nature-454-Walsh} was 
    suggested by the detailed 3-D shape of the NEO
    (66391) 1999 KW$_4$ \citep{2006-Science-314-Ostro}.

  \subsection{Disk-resolved imaging\label{ssec: hari}}
    \indent Since the 1990s, with the advent of the Hubble Space
    Telescope and large ground-based telescopes equipped with
    adaptive optics (AO: Lick, CFHT, Keck, ESO VLT, Gemini),
    we have access to the angular resolution required to resolve the
    apparent disk of asteroids
    \citep[\eg,][]{1993-Icarus-105-SaintPe2,
      1993-Icarus-105-SaintPe1, 
      1998-Icarus-132-Drummond, 
      2002-AJ-123-Parker,2006-AdSpR-38-Parker}.\\
    \indent From a time-series of disk-resolved images,
    spin-vector coordinates can be derived (using previous knowledge of the
    rotation period) 
    by analyzing the changes in the apparent shape of the asteroid
    profile \citep[see for instance][]{1997-Icarus-128-Thomas,
      2008-Icarus-197-Drummond}, or by following the apparent path 
    taken by an
    albedo patch on the surface during the rotation
    \citep[\eg,][]{2005-Nature-437-Thomas,
      2008-AA-478-Carry}.
    Triaxial shapes (ellipsoids) can also be derived
    \citep[see][for instance]{2009-Icarus-202-Drummond,
      2009-Science-326-Schmidt}, and 
    topography (such as the presence of facets or craters)
    studied from profile deviations to the ellipsoid
    \citep[\eg,][]{1997-Science-277-Thomas,
      2007-Icarus-191-Conrad}.
    With sufficient spatial resolution, imaging of the asteroid disk can
    allow construction of albedo maps of the surface,
    allowing the study of composition
    heterogeneity or localized space weathering effects
    \citep[\eg,][]{1997-Icarus-128-Binzel,
      2006-Icarus-182-Li, 2010-Icarus-208-Li,
      2008-AA-478-Carry, 2010-Icarus-205-Carry-a}. 
    In the case of asteroids visited by spacecraft,
    high spatial-resolution and precise photometry can be used to
    derive precise shape and digital terrain models
    by using stereophotoclinometric techniques
    \citep[see examples in][]{2008-MPS-43-Gaskell}.\\
    \indent The size and 3-D shape resulting from disk-resolved images are of
    great value, being obtained in a direct manner
    (as compared to
    an indirect determination of the shape from lightcurve inversion,
    for instance).
    The best angular resolution\,\footnote{limited
      by the diffraction, which acts as a
      low-pass filter with a \add{cutoff} frequency approximated by
      $\Theta$ = $\lambda$/$\mathcal{D}$ (radian), $\lambda$ \add{being}
      the wavelength and $\mathcal{D}$ the diameter of the telescope
      aperture.\label{foot: diff}} 
    of current Earth-based telescopes is
    about 0.04\arcsec. Due to systematics, however, we
    have found that our ability to accurately measure sizes
    degrades rapidly below about 0.10\arcsec, based on simulations and
    observations of the moons of Saturn and other objects 
    \citep{these-carry, 2009-DPS-41-Drummond}.  The sample of asteroids
    observable (\ie, having angular sizes that get above about 0.10\arcsec)
    is therefore limited to about 200 (over a given 10-year span).\\

  \subsection{Interferometry}
    \indent Apart from building larger telescopes, one efficient way
    to improve the angular resolution is to
    combine light beams from separated telescopes and
    \add{to} observe the resulting interference \add{(fringes)}.
    In such a mode,
    each telescope aperture plays the role of a sample aperture within
    a virtual telescope whose extent is the largest distance between
    the two telescopes (the spatial resolution $\Theta$ is still given by the
    equation in footnote $\ref{foot: diff}$, except that $\mathcal{D}$ is now the
    \textsl{distance} between the apertures).
    For instance, with telescopes separated by about 80\,m,
    the VLTI provides an order-of-magnitude improvement
    in angular resolution over a
    single telescope of the VLT (8\,m aperture). \\
    \indent This improvement in the resolution, however, occurs at the
    price of a loss in complete spatial
    information, because the
    virtual aperture is under-sampled.
    The very high angular-resolution is limited, at a given
    instant, to a single line
    \rem{in the sky plane,}
    \add{on the plane of the sky,}
    which is \add{given}
    by the baseline linking the two apertures. Along that baseline, the
    signal is directly related to the Fourier transform of the flux
    distribution on the plane of the sky.
    To expand the coverage in the spatial-frequency domain,
    and thus \add{to} allow the construction of 2-D images of the target, one
    must increase the number of projected baselines.
    This is
    commonly achieved by either adding multiple physical baselines
    (\ie, adding more telescopes) and/or by making observations throughout
    the night, as the Earth's rotation causes a progression of the
    position angle of the baseline on the sky.
    However, because asteroids typically complete their rotation
    in only a few hours,
    the method using multiple telescopes is more effective at
    providing higher spatial sampling. 
    Below we describe another promising
    technique to effectively increase the number of baselines, by using
    a Fizeau design.\\
    \indent For objects with no a priori information, the overall
    dimensions and spin properties
    can be determined under the assumption that the shape is
    well-described by a triaxial ellispoid
    \citep{2011-Icarus-211-Li}.
    For example, the two orthogonal directions of the Fine
    Guidance Sensor (FGS) on-board HST have been used to study binarity and
    measure the size of several MBAs
    \citep{2001-Icarus-153-Tanga, 2002-AA-391-Hestroffer,
      2003-AA-401-Tanga}.
%    Pushing further,
%    \citet{2006-IP-22-Kaasalainen} used FGS data
%    in combination with optical lightcurves to refine the 3-D shape
%    model of (15) Eunomia. \\
    \add{The fringes of interference, however, also contain information on
      the apparent shape, and can be used in combination with other
      data to derive 3-D shape models.
      For instance, \citet{2006-IP-22-Kaasalainen} used FGS data
      in combination with optical lightcurves to refine the 3-D shape
      model of (15) Eunomia from \citet{2002-Icarus-159-Kaasalainen}.} \\
    \indent Interferometry in the mid-infrared has been used recently also,
    with promising results.
    In this wavelength regime, the signal is linked to the distribution of
    temperature (\ie, \textsl{emitted} light) on the surface of the target
    (as opposed to the
    reflected light seen at visible wavelengths).
    \citet{2006-Icarus-181-Delbo} and
    \citet{2009-ApJ-694-Delbo}
    have combined thermal infrared TPM models
    (see Sect.~\ref{ssec: thermal}) with interferometry, allowing the size
    of 3-D shape models to be set
    \citep[similar to stellar occultation and
    thermal-infrared-only, see Sections~\ref{ssec: occ} and~\ref{ssec: thermal}
    and][]{2011-Icarus-215-Matter}.\\
    \indent The current limitation of interferometry is set by the
    sensitivity of available facilities
    and is driven largely by the
    integration time, which is usually short.
    Because the light beams from the two apertures traverse
    different paths in the atmosphere, their wavefronts encounter
    different turbulence-dominated perturbations, leading to a shift in
    their phase.
    Delay lines are used to ``slow down'' one beam with respect to the
    other, and \add{to} match their phase.
    The technical difficulties in maintaining coherence in this
    process limits the integration time to few 
    hundredths or thousandths of a second.
    A new generation of instruments (like PRIMA at VLT) with
    fringe-tracking systems will overcome this limitation in the near
    future, 
    \rem{pushing the technique toward fainter sources}
    \add{allowing fainter sources to be targeted}.\\
    \indent An interesting compromise between
    co-axial interferometers and traditional, filled aperture, telescopes
    can be found in Fizeau-type instruments
    such as LINC-NIRVANA, being built for 
    the Large Binocular Telescope \citep[LBT,
      see][]{2010-SPIE-7733-Hill},
    in which improved coverage of the \add{spatial-frequency plane
      (commonly called uv-plane, with u and v standing for the two
      orthogonal unit directions)}
      can be achieved in less
    telescope time. 
    This design (Fizeau \textsl{vs.} pupil-plane interferometer like
    the VLTI) allows instantaneous filling of the
    uv-plane up to the frequency set by the 8\,m apertures
    (Fizeau interferometers are true imaging devices and
    produce direct images of the plane of the sky), with
    an additional filling along one dimension up to the frequencies
    corresponding to the maximal baseline (22.7\,m for LBT). 
    Two or three epochs, with different position angles of the baseline
    on the plane of the sky, will be enough to fill the uv-plane up to
    the maximal extent of the telescope (achievable for transiting
    asteroids, especially with high elevation,
    when the position angle evolves quickly).
    \rem{The design of the LBT is therefore close to a 22.7\,m}
    \add{For this purpose, therefore, the LBT is equivalent to a 22.7\,m}
    telescope with a mask, corresponding to
    the configuration of component apertures, placed in the entrance
    pupil. \\
    \indent 
    The opening of 
    interferometric studies to longer wavelengths
    also has great potential, with 
    the use of millimeter and sub-millimeter arrays, where
    there are fewer technological limitations than in the optical
    range. 
    Future facilities such as ALMA, with 50 antennas
    \citep[translating into
    1\,225\,baselines \textsl{vs}. only 2\,baselines for MIDI at the VLTI,
    see][]{2009-ApJ-694-Delbo}, will allow dense spatial coverage,
    together with an angular resolution of few milli-arcseconds.
    Simulations have shown that several hundreds of MBAs and
    TNOs will be observable with ALMA
    \citep[see][]{2009-Icarus-200-Busch, 2011-Icarus-213-Moullet}.
    Interferometry at thermal wavelengths (mid-infrared to millimeter),
    in combination with other techniques (\eg, lightcurves,
    see Sect.~\ref{ssec: lc}), will thus allow the
    derivation of 3-D shape models, together with thermal
    properties, for many asteroids of small apparent diameter.

\section{The KOALA algorithm\label{sec: koala}}
  \indent With advantages and drawbacks of each observing technique in mind, we
  have developped a multi-data inversion algorithm: Knitted
  Occultation, Adaptive-optics, and Lightcurve Analysis (KOALA), that
  makes simultaneous use of data from three distinct observation
  types\,\footnote{We categorize dense (\ie, lightcurves) and sparse photometry
    together to form a single data type.}
  to determine the physical properties of asteroids
  \citep{2010-Icarus-205-Carry-a, 2011-IPI-5-Kaasalainen}.
  KOALA takes advantage of the direct measure of the
  apparent size and shape of asteroids on the plane of the sky
  provided by the timings of stellar occultations and disk-resolved
  images, and of the indirect constraints on spin and 3-D shape
  given by lightcurves.
  \add{We quickly summarize below how the inversion works
    \citep[see][for a comprehensive description of the
      algorithm]{2011-IPI-5-Kaasalainen}.} \\ 
  \indent KOALA is a step-iterative minimization algorithm,
  solving for the spin parameters
  (spin-vector coordinates $\lambda$, $\beta$, and sidereal period
  $\mathcal{P}$),
  3-D shape
  (given by a set of $\mathcal{N}$ coefficients of spherical harmonics,
  including the overall size),
  phase function
  \citep[defined as a three-parameter model, see][]{
    2001-Icarus-153-Kaasalainen-a, 2001-Icarus-153-Kaasalainen-b}, and
  scattering law
  \citep[generally taken as a combination of the
    Lommel-Seeliger and Lambert diffusion laws,
    following][although other models such as Hapke can be
    used]{2001-Icarus-153-Kaasalainen-a}.\\ 
  \indent From a set of estimated parameters (determined from 
  lightcurve-only inversion or analysis of the disk-resolved images,
  for instance),
  a trial solution is created, with associated synthetic
  data sets (\ie, simulated lightcurves, 
  disk-resolved images, and occultation profiles).
  KOALA then follows a Levenberg-Marquardt minimization scheme to
  determine the set of parameters that best fit all the data sets
  simultaneously, by comparing at each step the synthetic
  data with real measurements.
  The iteration stops when the residuals between the
  simulated and measured data sets reach an acceptable level
  (\ie, the level of the intrinsic noise of the measurements). \\
  \indent As a safeguard, the resolution (\ie,
  $\mathcal{N}$) is set to the lowest possible value for
  which a fit to the data can be achieved.
  We also introduced several regularizations:
  (a) a non-convexity (``smoothness'') term, that avoids 
  spurious features (unrealistic 
  topography) at small scales, unconstrained by the data;
  and 
  (b) an inertia tensor regularization for principal-axis rotators,
  that forces the asteroid spin-axis to remain aligned, within a
  few degrees, with the largest moment of inertia.
  The relative weights of the different data types and
  also of the regularizations are determined using the maximum compatibility
  estimate of \citet{2011-IPI-5-Kaasalainen}, instead of being
  subjective.

\section{KOALA and Rosetta-flyby shape model\label{sec: comp}}
\begin{figure}[!t]
  \includegraphics[width=.5\textwidth]{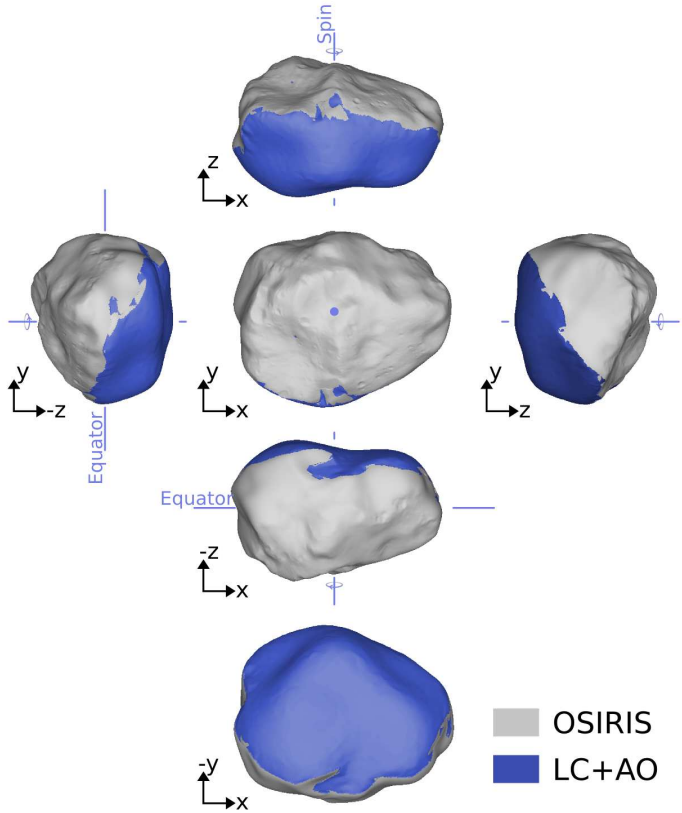}
  \caption{%
    \add{Lutetia 3-D shape model from \citet{2011-Science-334-Sierks}, 
    displayed in a net layout.
    Regions imaged by OSIRIS are displayed in gray. The part of the
    shape model that is based on ground-based data only
    (lightcurves and images: LC+AO) is plotted in blue. }
    \label{fig: LAM}
    }
\end{figure}

  \indent On 2010 July 10, the ESA Rosetta 
  spacecraft made a close flyby of
  the main-belt asteroid (21) Lutetia.
  In support of the mission, we had combined 
  optical lightcurves with disk-resolved images,
  from ground-based AO. We produced
  a full 3-D shape model of Lutetia, using KOALA,
  months before the encounter
  \citep[see][and Fig.~\ref{fig: models}]{2010-AA-523-Drummond,
    2010-AA-523-Carry}. 
  This flyby provided a rare opportunity to
  test and calibrate KOALA with close-up spacecraft imaging.\\
  \indent The closest approach (CA) occurred at 3\,170\,km from the asteroid
  at a relative speed of 15\,km/s.
  The Narrow Angle Camera (NAC)
  of the OSIRIS instrument on-board Rosetta
  \citep{2007-SSRv-128-Keller} 
  returned a total of 
  202\,images
  during the
  flyby, which spanned about 9\,h. 
  The NAC image scale ranged from 5\,000 to 60\,m/pix,
  reaching its minimum value at CA. 
  These very high spatial-resolution images
  have been used to produce detailed
  3-D shape models of Lutetia, using stereophotoclinometry
  \citep{2011-Science-334-Sierks}, and
  stereophotogrammetry \citep{2011-PSS--Preusker}.
  We only consider here the model from \citet{2011-Science-334-Sierks},
  these two models being similar enough at the medium-to-large
  scale for our purpose.\\
%
%
%%%%%%%%%%%%%%%%%%%%%%%%%%%%%%%%%%%%%%%%%%%%%%%%%%%%%%%%%%%%%%%%%%%%%%%%%%%%%%%%%%%
%%%%%%%%%%%%%%%%%%%%%%%%%%%%%%%%%%%%%%%%%%%%%%%%%%%%%%%%%%%%%%%%%%%%%%%%%%%%%%%%%%
\begin{table}
\begin{tabular}{l@{}c|rc@{$\pm$}cc@{$\pm$}cc@{$\pm$}c|l@{ }l@{ $\pm$ }l}
& & \multicolumn{7}{c|}{Diameters (km)} &
    \multicolumn{3}{c}{Spin axis (\degr)} \\
& & \multicolumn{1}{c}{$d$} & 
    \multicolumn{2}{c}{$a$} &
    \multicolumn{2}{c}{$b$} & 
    \multicolumn{2}{c|}{$c$} &
    $\lambda$ &
    $\beta$ & 
    \multicolumn{1}{c}{$\sigma$} \\
 \hline
KOALA & \includegraphics[width=1em]{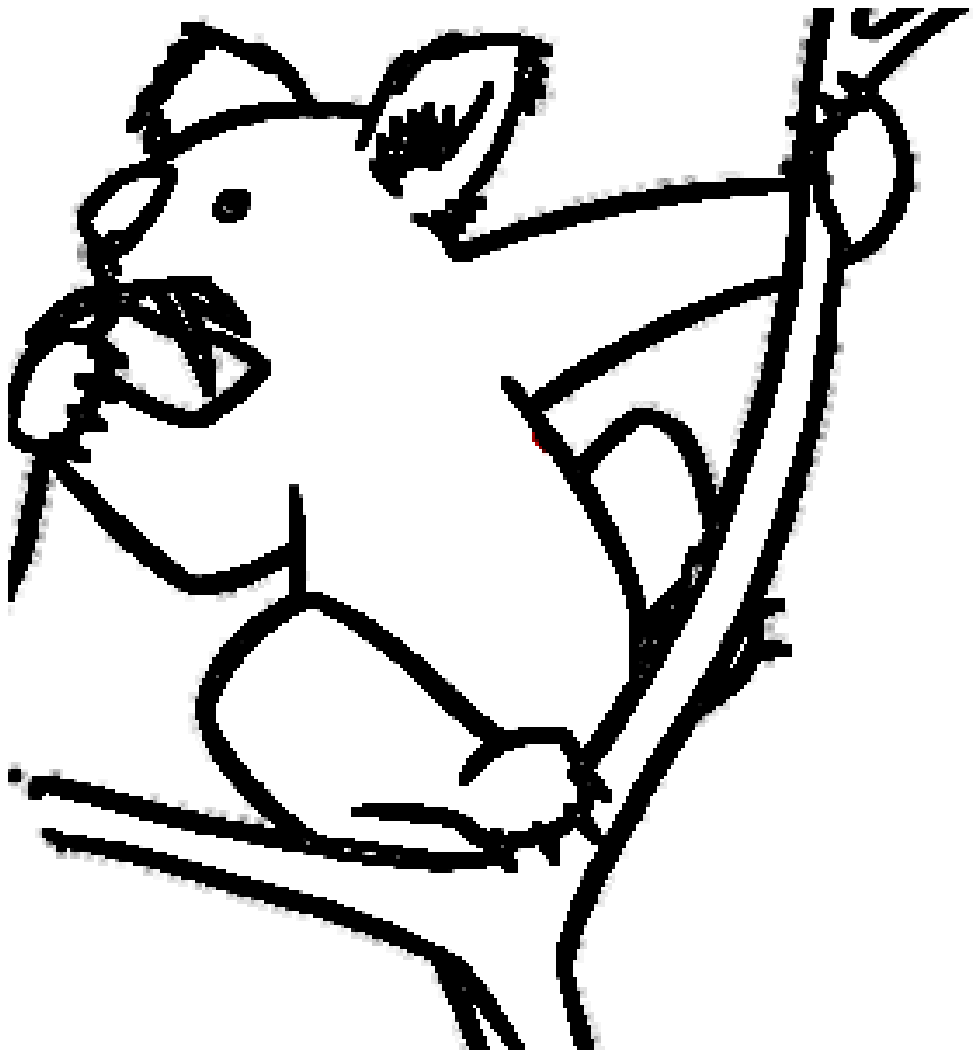}  & 105 & 124 & 5 & 101 & 4 & 93 & 13 & 52   & -6   &  5  \\
OSIRIS& \includegraphics[width=1em]{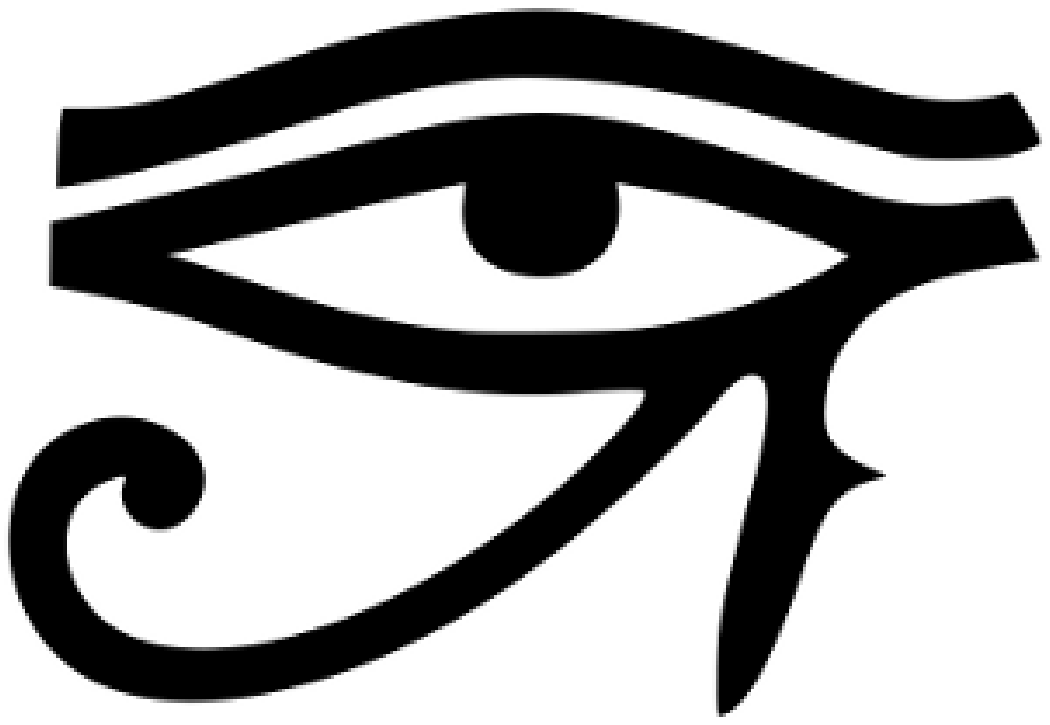} &  98 & 121 & 1 & 101 & 1 & 75 & 13 & 52.2 & -7.8 & 0.4 \\
 \hline
\end{tabular}
\caption{%
  Comparison of the dimensions and spin-vector coordinates of (21) Lutetia
  derived from OSIRIS
  images during the ESA Rosetta flyby
  \citep{2011-Science-334-Sierks} with
  those derived by KOALA
  \citep{2010-AA-523-Drummond,2010-AA-523-Carry} prior to the flyby.
  Volume-equivalent diameter ($d$) and
  triaxial diameters ($a > b >c$) are reported in km,
  and spin-vector coordinates in degree
  (longitude $\lambda$, latitude $\beta$ in the ECJ2000 reference
  frame, with $\sigma$ the angular radius of the uncertainty circle).
\label{tab: abc}
}
\end{table}
%%%%%%%%%%%%%%%%%%%%%%%%%%%%%%%%%%%%%%%%%%%%%%%%%%%%%%%%%%%%%%%%%%%%%%%%%%%%%%%%%%
%%%%%%%%%%%%%%%%%%%%%%%%%%%%%%%%%%%%%%%%%%%%%%%%%%%%%%%%%%%%%%%%%%%%%%%%%%%%%%%%%%
%
%
  \indent \add{Lutetia's spin axis is tilted such that its pole is
    nearly in its orbital plane
    \citep[obliquity of 96\degr, see][]{2010-AA-523-Carry,
      2011-Science-334-Sierks}.
  At the time of the Rosetta flyby, the southern hemisphere was in
  seasonal shadow, and observations at optical/\add{near-infrared} wavelengths
  were not possible south of -40\degr\,latitude.
  The detailed 3-D shape model derived from flyby images 
  by stereophotoclinometry \citep{2011-Science-334-Sierks}
  therefore does not cover a large fraction of the asteroid's
  southern hemisphere (see Fig.~\ref{fig: LAM}),
  the southernmost portion of the shape model
  being determined using the KOALA algorithm together with the
  ground-based data from 
  \citet{2010-AA-523-Carry}.
  Almost all of the AO images that entered
  into the KOALA solution, even though from multiple epochs, were
  taken looking at either high southerly or high northerly
  sub-Earth latitudes
  \citep[see][]{2010-AA-523-Carry, 2010-AA-523-Drummond}.
  For both the ground-based KOALA and the flyby analyses,
  therefore, the shortest ($c$) dimension (Table~\ref{tab: abc})
  is not well
  constrained and represents the largest contribution to the
  uncertainty in the volume estimates.}\\
%-TBD
  \indent \rem{An additional observation was
  taken with a more equatorial geometry, and presented by
  \citet{2010-AA-523-Drummond}.
  The purpose of this observation was a search for satellites
  and therefore the Point-Spread Function (PSF) calibrations were not
  adequate for shape recovery.
  With their method
  based on the hypothesis that the shape is well-described by an
  ellipsoid, \citeauthor{2010-AA-523-Drummond} could use this
  observation while it was not used by neither
  \citeauthor{2010-AA-523-Carry} nor
  \citeauthor{2011-Science-334-Sierks}.  
  This observation gave us our primary leverage to constrain the size of
  the KOALA
  shape model along its short axis ($c$ dimension) to be 93\,km rather
  than the 80\,km derived without it
  \citep[see details in][]{2010-AA-523-Drummond,
    2010-AA-523-Carry}.\\ }
  \indent
  We first present (Section~\ref{ssec: abc})
  an overall comparison of our size and spin estimates
  (based on ground-based techniques)
  with those derived from the analysis of
  the very high spatial-resolution images acquired by
  OSIRIS \citep{2011-Science-334-Sierks}.
  We then directly compare the predicted shape profiles, based on KOALA,
  with those images (Section~\ref{ssec: kvo}), seeking to calibrate our method.

%
%%%%%%%%%%%%%%%%%%%%%%%%%%%%%%%%%%%%%%%%%%%%%%%%%%%%%%%%%%%%%%%%%%%%%%%%%%%%%%%%%%
%%%%%%%%%%%%%%%%%%%%%%%%%%%%%%%%%%%%%%%%%%%%%%%%%%%%%%%%%%%%%%%%%%%%%%%%%%%%%%%%%%
%%%%%%%%%%%%%%%%%%%%%%%%%%%%%%%%%%%%%%%%%%%%%%%%%%%%%%%%%%%%%%%%%%%%%%%%%%%%%%%%%%
%%%%%%%%%%%%%%%%%%%%%%%%%%%%%%%%%%%%%%%%%%%%%%%%%%%%%%%%%%%%%%%%%%%%%%%%%%%%%%%%%%
%
%
%
%%%%%%%%%%%%%%%%%%%%%%%%%%%%%%%%%%%%%%%%%%%%%%%%%%%%%%%%%%%%%%%%%%%%%%%%%%%%%%%%%%
%%%%%%%%%%%%%%%%%%%%%%%%%%%%%%%%%%%%%%%%%%%%%%%%%%%%%%%%%%%%%%%%%%%%%%%%%%%%%%%%%%
\begin{figure}[!t]
  \includegraphics[width=.5\textwidth]{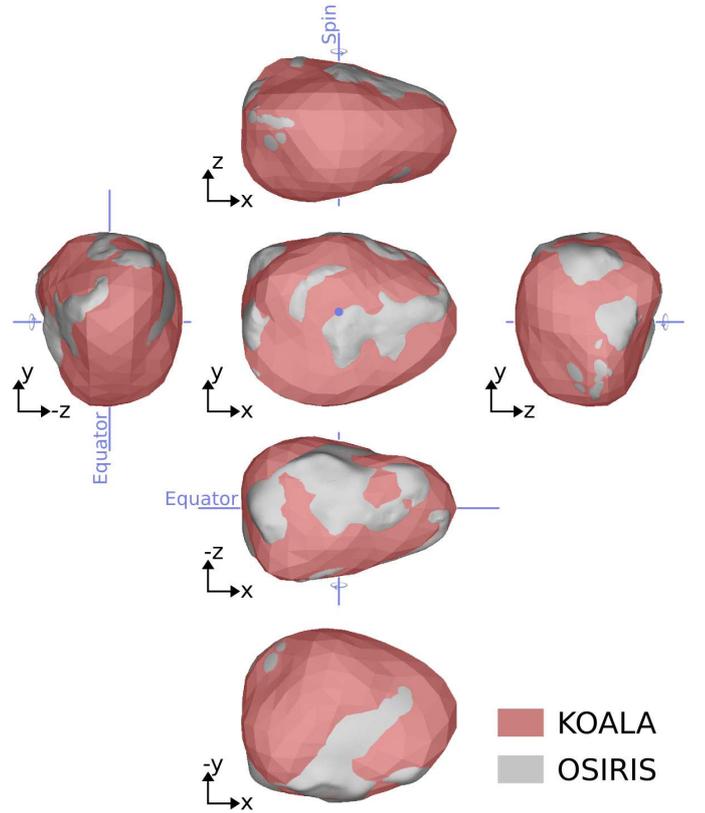}
  \caption{%
    \add{Similar plot as Fig.~\ref{fig: LAM},
      showing the agreement
      between the pre-flyby KOALA
      \citep[reddish,][]{2010-AA-523-Carry} 
      and post-flyby
      \citep[grayish,][]{2011-Science-334-Sierks} shape models of
      (21) Lutetia.}
%    OSIRIS \citep[grayish,][]{2011-Science-334-Sierks} and
%    KOALA \citep[reddish,][]{2010-AA-523-Carry} shape models of
%    (21) Lutetia plotted together
%    (displayed in a net layout).
    The radii of the KOALA model are larger in red regions, and
    those of the OSIRIS model in gray regions.
    \rem{This}
    \add{The amalgam of gray and red colors}
    illustrates the spectacular agreement between
    the results from \add{the} Rosetta flyby and our KOALA
    model obtained before the flyby.
    The dominance of red hues, however, highlights the
    relativly lower sensitivity to
    concavities of the KOALA model with respect
    to the OSIRIS 3-D shape.
    Shown in this comparison is the original KOALA model of
    \citet{2010-AA-523-Carry}, with a $c$-dimension of
    80\,km (overall size 124\,$\times$\,101\,$\times$\,80\,km),
    rather than the modified KOALA (``hybrid'')
    estimate of \citet{2010-AA-523-Drummond}, having
    a $c$-dimension of 93\,km and shown in Table~\ref{tab: abc}.
    This is a more realistic
    comparison for the figure because much of the southern hemisphere
    of the OSIRIS model \add{(Fig.~\ref{fig: LAM})}
      is based on the \add{data that entered into the}
    original KOALA model.
    \label{fig: net}
    }
\end{figure}
%\clearpage
%%%%%%%%%%%%%%%%%%%%%%%%%%%%%%%%%%%%%%%%%%%%%%%%%%%%%%%%%%%%%%%%%%%%%%%%%%%%%%%%%%
%%%%%%%%%%%%%%%%%%%%%%%%%%%%%%%%%%%%%%%%%%%%%%%%%%%%%%%%%%%%%%%%%%%%%%%%%%%%%%%%%%
%
%

  \subsection{Overall comparison\label{ssec: abc}}
\begin{figure}[!t]
  \includegraphics[width=.5\textwidth]{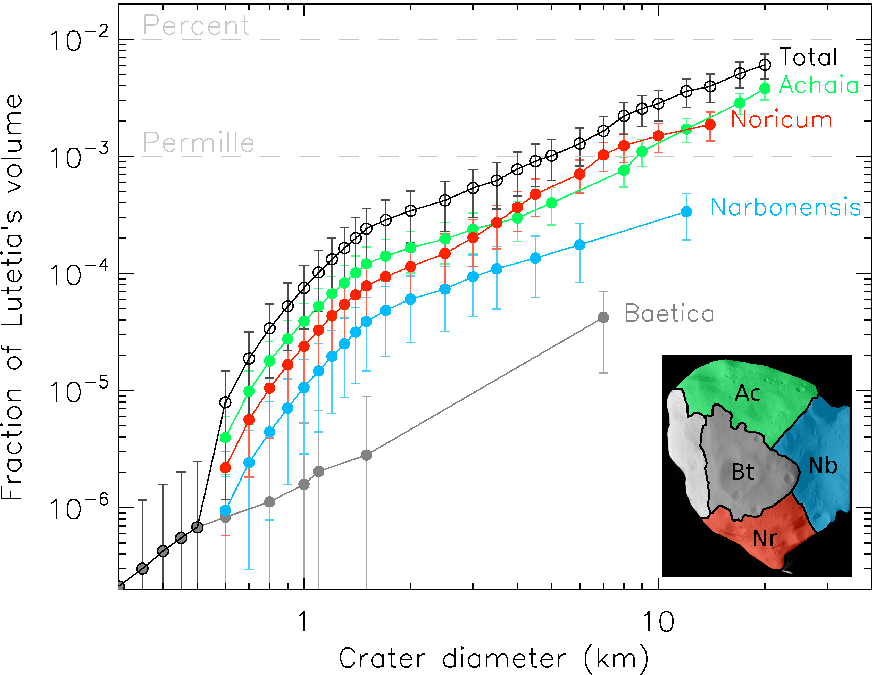}
  \caption{%
    \add{Cumulative distribution of the volume encompassed by
      craters as function of their diameter, for four
      geomorphological units of Lutetia: 
      Achaia (Ac), Baetica (Bt), Narbonensis (Nb), and Noricum (Nr)
      (see the insert, adapted from Fig.~1 by
      \citet{2011-Science-334-Sierks}, 
      and also \citet{2011-PSS--Thomas} for the definition of these units).}
    \label{fig: crat} }
\end{figure}

    \indent In this section, we present a comparison of the
    spin-vector coordinates and dimensions derived from the flyby
    with the pre-flyby values from KOALA (Table~\ref{tab: abc}).
    Because southerly latitudes were not visible during the flyby
    and the $c$-dimension in the OSIRIS model comes from
    a combination of OSIRIS imaging and KOALA model information,
    the OSIRIS $c$-dimension is not wholly independent.
    But the spin-vector coordinates and equatorial dimensions ($a$, $b$)
    from the OSIRIS analysis can be used as ``ground-truth'' to
    calibrate our KOALA method.\\ 
    \indent
    First, the spin-vector coordinates agree to within two
    degrees\,\footnote{\add{the coordinates obtained with KOALA
      are 1.2\degr~and 1.8\degr~from
      the spin-vector solutions derived
      by F. Preusker ($\lambda$\,=\,52.6\degr, $\beta$\,=\,-7.1\degr, pers. communication)
      and
      by \citet{2011-Science-334-Sierks} respectively,
      using the OSIRIS images.}}, 
    well inside the 5\degr~uncertainty quoted for KOALA
    Then, equatorial dimensions ($a$ and $b$) are within 3\,km of the
    flyby estimates,
    again well within the 
    uncertainties reported for each dimension using KOALA.
%--TBD
    \add{The larger difference between the estimates of the
    short ($c$) axis results from the expansion of the KOALA
    $c$ dimension from its original 80\,km to 93\,km,
    based on the analysis by
    \citet{2010-AA-523-Drummond}.
    Indeed,
    the best-fit solution for all the AO images
    (including an additional observation 
    taken with a more equatorial geometry, presented by
    \citet{2010-AA-523-Drummond} and not used by
    \citet{2010-AA-523-Carry} owing to calibration issues, see 
    \citeauthor{2010-AA-523-Carry} for details)
    pointed toward a larger $c$ dimension
    \citep[supported by independent considerations on the amplitude of
      lightcurves by][]{2010-AA-515-Belskaya}.}
    The volume-equivalent diameter is, therefore, larger
    (105 \textsl{vs}. 98\,km)
    for the KOALA
    model than for the flyby-derived model by 
    \citet{2011-Science-334-Sierks}.
    Disk-resolved images of Lutetia taken with low
    sub-observer latitudes are required 
    to confirm its $c$ dimension,
    and \add{to} set tighter constraints on its volume.\\
    \indent \add{The depth of large-scale concavities was slightly
    underestimated by KOALA (Fig.~\ref{fig: net});
    estimating the depth of a concavity from profiles only is
    problematical because the concavity is
    hidden behind its rim.
    With stereophotoclinometry, is is possible to sense depths.
    In addition, while KOALA is sensitive to large-scale concavities,
    the necessarily limited 
    resolution when imaging from a distance of 200\,million km means that
    it is less sensitive to small-to-medium-scale concavities.
    We evaluated the influence of craters on the volume of Lutetia.
    For that, we used the crater size distribution measured by
    \citet{2011-PSS--Marchi}, and estimated their volume as that of a
    spherical cap ($V_{\rm crat}$),
    using the average depth-to-diameter ratio measured by 
    \citet{2011-PSS--Vincent}:}
\begin{equation}
  V_{\rm crat} = \frac{\pi}{6} d \left( \frac{3}{4} D^2 + d^2 \right)
\end{equation}
    \noindent \add{where $d$ is the depth of the crater and $D$ its
      diameter. We present in Fig.~\ref{fig: crat} the cumulative
      distribution of the volume encompassed by craters, counted on four
      geomorphological units
      \citep[see][, for a detailed definition of the
        units]{2011-PSS--Thomas}, against their diameter. 
      The vast majority of the volume is due to the handful of craters
      with diameters between 10 and 20\,km.
      The total influence of these craters 
      on the volume of Lutetia is 0.6\,$\pm$\,0.1\%,
      \ie, the volume of Lutetia would be 0.6\% larger if these craters
      were not included in the 3-D shape model.
      Extrapolating this value, by considering the area covered by these
      unit to the whole surface of Lutetia, make the
      total\,\footnote{\add{Because the shape of the southern
          hemisphere is poorly constrained,
        and the number of craters there unknown,
        this extrapolated value is only a rough estimate.}}
      influence of craters to be 2.4\,$\pm$\,0.6\%.}\\
    \indent \add{Proper modeling of the craters in the 3-D shape model
      \citep[from high-resolution flyby images,][]{2011-Science-334-Sierks}
      is therefore crucial, given the level of accuracy reached
      elsewhere on the surface.
      For Earth-based observations, however, this will remain a minor
      source of uncertainty:
      depending on the methods, 
      the volume accuracy ranges from few percent to few tens of
      percent (see Sect.~\ref{sec: intro}).
      We discuss in the next section (\ref{ssec: kvo}) the influence
      of these craters on the volume accuracy that can be reached
      using KOALA.}

%%%%%%%%%%%%%%%%%%%%%%%%%%%%%%%%%%%%%%%%%%%%%%%%%%%%%%%%%%%%%%%%%%%%%%%%%%%%%%%%%%
%%%%%%%%%%%%%%%%%%%%%%%%%%%%%%%%%%%%%%%%%%%%%%%%%%%%%%%%%%%%%%%%%%%%%%%%%%%%%%%%%%
%\clearpage
%%%%%%%%%%%%%%%%%%%%%%%%%%%%%%%%%%%%%%%%%%%%%%%%%%%%%%%%%%%%%%%%%%%%%%%%%%%%%%%%%%
%%%%%%%%%%%%%%%%%%%%%%%%%%%%%%%%%%%%%%%%%%%%%%%%%%%%%%%%%%%%%%%%%%%%%%%%%%%%%%%%%%
%

  \subsection{Detailed analysis\label{ssec: kvo}}
\begin{figure}
  \includegraphics[width=.5\textwidth]{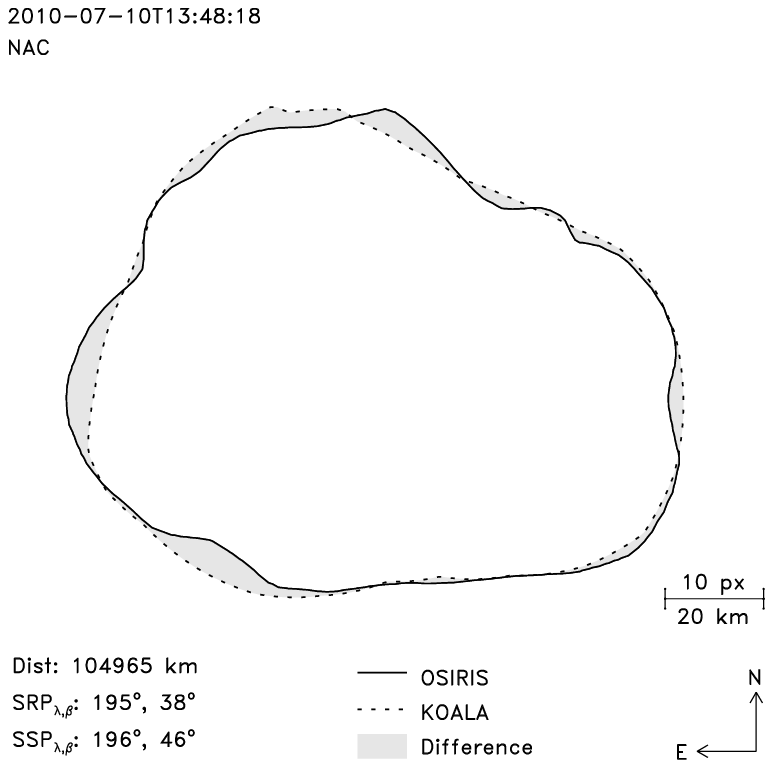}
  \caption{%
    Example of a profile comparison, as measured on the OSIRIS NAC detector
    plane (solid line) and simulated from the KOALA model (dotted
    line), taken at 13:48:18 UT (CA - 1h57m).
    The light gray area represents the difference in projected area
    on the plane of the sky between the prediction and the observation.
    We report the Rosetta-Lutetia distance, the coordinates of the
    sub-Rosetta point (SRP) and subsolar point (SSP), and a scale for
    angular \add{(OSIRIS NAC pixels)} and physical dimension.
    \label{fig: profiles}
    }
\end{figure}

\begin{figure*}
  \includegraphics[width=\textwidth]{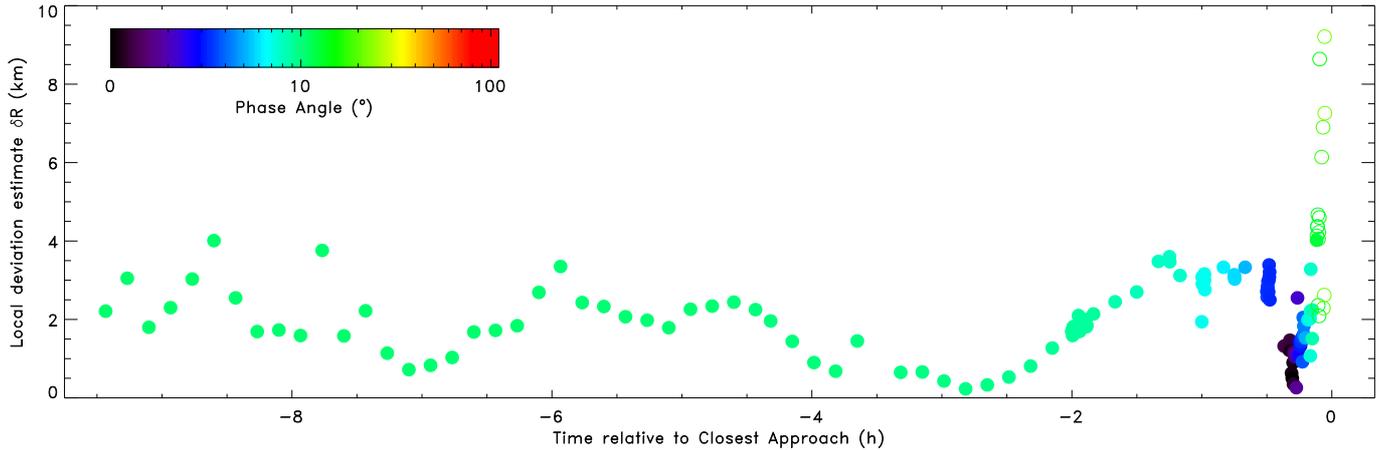}
  \caption{%
    Estimate of the local deviation ($\delta R$) between the profiles of Lutetia
    extracted from the OSIRIS images and the prediction from the KOALA
    model \citep{2010-AA-523-Carry}.
    The symbols are color-coded as a function of the phase angle.
    Filled symbols correspond to images where the phase angle was
    smaller than 10\degr~(see text). 
    \label{fig: local}
  }
\end{figure*}

    \indent We push further the calibration of the KOALA method by
    comparing comprehensively the KOALA shape model
    predictions with the very high-spatial-resolution images provided
    by OSIRIS NAC. 
    In the absence of a complete 3-D shape model, based
    on an entirely independent
    data set, it is difficult to fully calibrate the
    volume estimate provided by KOALA.
    For each OSIRIS image, we extracted the profile of the apparent
    disk of Lutetia, composed of its limb
    and its terminator.
    We produced synthetic views of the KOALA shape model under the
    same geometry (\ie, as seen from Rosetta, see
    Fig.~\ref{fig: profiles}): phase angle, subsolar
    point (SSP) 
    and sub-Rosetta point (SRP) coordinates, using the 
    Miriade\,\footnote{\href{http://vo.imcce.fr/webservices/miriade/}{http://vo.imcce.fr/webservices/miriade/}}    
    VO ephemeris
    generator\,\footnote{we used Rosetta flight
    kernel ORHR$\_\_\_\_\_\_\_\_\_\_\_\_\_\_\_$00122.BSP}.\\
%    \add{\citep{2008-ACM-Berthier}}.\\
%
%
%
%
%%%%%%%%%%%%%%%%%%%%%%%%%%%%%%%%%%%%%%%%%%%%%%%%%%%%%%%%%%%%%%%%%%%%%%%%%%%%%%%%%%
%%%%%%%%%%%%%%%%%%%%%%%%%%%%%%%%%%%%%%%%%%%%%%%%%%%%%%%%%%%%%%%%%%%%%%%%%%%%%%%%%%
%%%%%%%%%%%%%%%%%%%%%%%%%%%%%%%%%%%%%%%%%%%%%%%%%%%%%%%%%%%%%%%%%%%%%%%%%%%%%%%%%%
%%%%%%%%%%%%%%%%%%%%%%%%%%%%%%%%%%%%%%%%%%%%%%%%%%%%%%%%%%%%%%%%%%%%%%%%%%%%%%%%%%
%
%
%
%
    \indent We estimate the fit of KOALA predictions
    to OSIRIS data by computing the difference between the projected areas on
    the plane of the sky. The relative accuracy of
    the volume determination ($\delta V/V$) can then be determined as follows:
    \begin{equation}
      \frac{\delta V}{V} = \frac{3}{2} \frac{\mathcal{A}_O - \mathcal{A}_K}{\mathcal{A}_O}
      \label{eq: vol}
    \end{equation}

    \noindent where $\mathcal{A}_K$ and $\mathcal{A}_O$ are the
    areas of the KOALA prediction and on OSIRIS frame, respectively.
    Negative and positive $\delta V$ respectively indicate an
    overestimate or underestimate of the volume by the KOALA model.    
    We discarded from the current analysis the images taken close
    to CA, which have substantial phase angle (above 10\degr),
    where local
    topography (\eg, crater rims) produce large projected shadows,
    increasing the uncertainty in the prediction of the terminator
    position on the surface of Lutetia.\\
%
%
%%%%%%%%%%%%%%%%%%%%%%%%%%%%%%%%%%%%%%%%%%%%%%%%%%%%%%%%%%%%%%%%%%%%%%%%%%%%%%%%%%
%%%%%%%%%%%%%%%%%%%%%%%%%%%%%%%%%%%%%%%%%%%%%%%%%%%%%%%%%%%%%%%%%%%%%%%%%%%%%%%%%%
\begin{figure*}
  \includegraphics[width=\textwidth]{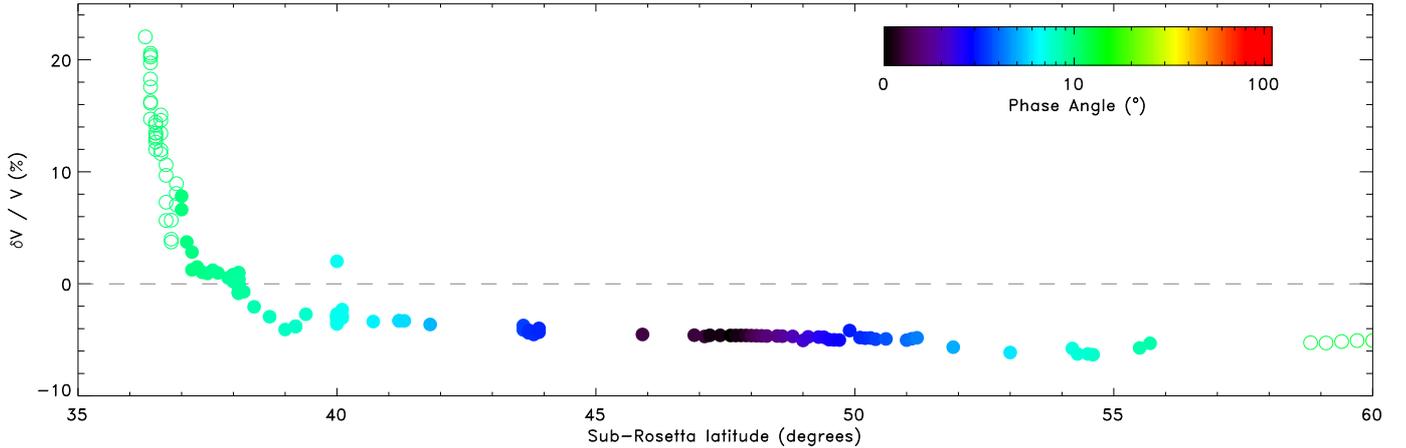}
  \caption{%
    Difference in volume estimated (Eq.~\ref{eq: vol})
    from the difference in projected
    area between the profiles of Lutetia
    extracted from the OSIRIS images and the prediction from the KOALA
    model \citep{2010-AA-523-Carry}.
    The symbols are color-coded as a function of the phase angle.
    Filled symbols correspond to images where the phase angle was
    smaller than 10\degr.
    \label{fig: vol}
    }
\end{figure*}
%%%%%%%%%%%%%%%%%%%%%%%%%%%%%%%%%%%%%%%%%%%%%%%%%%%%%%%%%%%%%%%%%%%%%%%%%%%%%%%%%%
%%%%%%%%%%%%%%%%%%%%%%%%%%%%%%%%%%%%%%%%%%%%%%%%%%%%%%%%%%%%%%%%%%%%%%%%%%%%%%%%%%
%
%
    \indent From 
    this detailed analysis, we confirm the results from the
    overall comparison presented in Section~\ref{ssec: abc}:
    the volume was slightly overestimated
    with KOALA, \rem{with respect} \add{relative}
    to the results derived from the flyby.
    But, as visible in Fig.~\ref{fig: vol}, the difference between
    the KOALA predictions and the OSIRIS images is
    almost constant, at about -5\%, for
    all images taken with phase angle smaller than $\sim$10\degr.
    The largest deviations, still within the uncertainty reported
    using KOALA, are found for sub-Rosetta latitudes
    lower than about 40\degr, \ie, for geometries that had not been
    observed from Earth with AO.
    Owing to the restricted geometries of ground-based AO imaging
    observations to date (always close to ``pole-on'', see 
    \citet{2010-AA-523-Drummond} and 
    \citet{2010-AA-523-Carry}),
    the differences between KOALA predictions and OSIRIS frames may be
    related to two distinct factors:
    one ``inherent'' to KOALA (which is the one we
    seek to evaluate), and another related to the observing geometry. 
    Unfortunately, there is no easy way to distinguish between
    these two effects.
    We thus estimate that the inherent uncertainty in the KOALA volume 
    is about 5\% (Fig.~\ref{fig: vol}).
    This uncertainty can increase, however, due to unfavorable
    observing geometries, such as we have for the present case of Lutetia.\\
    \indent In addition to calibrating the
    volume estimate,
    we assess the accuracy of the KOALA 3-D shape determination.
    From the direct comparison of both models (pre- and post-flyby,
    in Fig.~\ref{fig: net}), we can already qualitatively
    assert that KOALA allows accurate 3-D shape determination.
    However, many different shapes can result in similar volume,
    and also in similar overall triaxial-ellipsoid dimensions
    (which are accurate to a couple of km, see
    Section~\ref{ssec: abc}).
    We therefore apply the following criterion
    ($\delta R$, see Eq.~\ref{eq: radius}) to estimate the
    \textsl{local} deviation 
    of the KOALA model to the real shape of Lutetia, and therefore
    calibrate quantitatively the KOALA 3-D shape determination:
    \begin{equation}
      \delta R = \frac{1}{N} \sum_i^N \frac{(\mathcal{O}_i-\mathcal{K}_i)^2}{\mathcal{O}_i}
      \label{eq: radius}
    \end{equation}

    \noindent where $\mathcal{O}_i$ and $\mathcal{K}_i$, respectively, are the
    OSIRIS and KOALA profile radii of the $i$th point
    (out of $N$ describing the OSIRIS profile),
    measured from an arbitrary center. \\
    \indent We present in Fig.~\ref{fig: local}
    the estimate of the local deviation of the KOALA
    predictions to the apparent shape of Lutetia measured on the
    OSIRIS NAC images, as a function of the time relative to CA.
    The deviations are limited to about 4\,km at maximum, and are 
    about 2\,km on average
    \citep[confirmed by the independent analysis of][]{2011-PSS--Preusker}.
    The typical accuracy in the elevation of each vertex is therefore
    close to 2\,km. This is consistent with the overall
    comparison presented in Section~\ref{ssec: abc}, but we show here
    that not only the overall sizes are accurate to about 2\,km, but
    that this accuracy is maintained at local scales.\\
    \indent \add{The \textsl{apparent}
      absence of craters in the KOALA
      model\,\footnote{\add{Although the model does not show round,
          crater-like, features, KOALA allows modeling of concavities, that
      can be either large impact craters.}}
      \citep[compared to 3-D shape models derived from
        flyby, \eg,][]{2011-Science-334-Sierks, 2011-PSS--Preusker}
      has therefore little influence on the volume estimate and its
      accuracy. 
    The largest craters seen on Lutetia have a diameter of
    about 20\,km and a depth of about 2--3\,km
    \citep[see Fig.~\ref{fig: crat}, and][]{2011-PSS--Marchi,
      2011-PSS--Vincent},
    corresponding to the typical accuracy on elevation, 
    and are therefore already included in the 
    uncertainty envelop around the KOALA shape model.} 
%

%%%%%%%%%%%%%%%%%%%%%%%%%%%%%%%%%%%%%%%%%%%%%%%%%%%%%%%%%%%%%%%%%%%%%%%%%%%%%%%%%%
%%%%%%%%%%%%%%%%%%%%%%%%%%%%%%%%%%%%%%%%%%%%%%%%%%%%%%%%%%%%%%%%%%%%%%%%%%%%%%%%%%
%%%%%%%%%%%%%%%%%%%%%%%%%%%%%%%%%%%%%%%%%%%%%%%%%%%%%%%%%%%%%%%%%%%%%%%%%%%%%%%%%%
%%%%%%%%%%%%%%%%%%%%%%%%%%%%%%%%%%%%%%%%%%%%%%%%%%%%%%%%%%%%%%%%%%%%%%%%%%%%%%%%%%
%
%
    \indent From this detailed comparison of OSIRIS frames with KOALA
    predictions, we assert that 
    the KOALA model performed extraordinarily well, especially given
    two limiting factors related to the observing geometries of the
    ground-based observations. 
    First, for the imaging observations, Lutetia had an angular diameter of
    about 0.1\arcsec, close to the angular resolution for which we
    currently can extract useful shape information from ground-based
    telescopes with AO. 
    Second, for the oppositions in 2007 and 2008, Lutetia was positioned
    at diametrically opposite apparent ecliptic coordinates from
    Earth, and 
    because of the high obliquity, we observed Lutetia close to North-pole-on and
    then close to South-pole-on, and we were not able to achieve a good
    equatorial view, resulting in a poorly determined $c$-axis
    \citep{2010-AA-523-Drummond, 2010-AA-523-Carry}.
    We expect to improve that in upcoming observations.\\
    \indent We have determined \rem{that}
    the difference between the KOALA shape
    model and the \textsl{real} topography of Lutetia
    to be 2\,km, on average.
    Considering Lutetia's volume-equivalent diameter
      (98\,km, see Table~\ref{tab: abc}), this
    translates into a relative precision on the radii of about 5\%.
    We can therefore expect a conservative upper limit of
    15\% for the accuracy that can be achieved
    on volume estimates made by KOALA
    (including possible systematic effects), although it appears that the
    measured uncertainty (Fig.~\ref{fig: vol}) is closer to 5\%.
    These estimates of the accuracy achievable with KOALA are close to
    our previously estimated uncertainty:
    Although formal uncertainties for our
    shape-fitting algorithms are closer to 1 or
    2\,km, we had estimated that our size measurements were 
    affected by systematic errors at the level of 1--3\% 
    \citep[from simulations and observations of Saturn's moons,
      see][]{these-carry, 2009-DPS-41-Drummond}.\\

\section{KOALA and thermal radiometry\label{sec: thermal}}
\begin{figure}
  \includegraphics[angle=90, width=.5\textwidth]{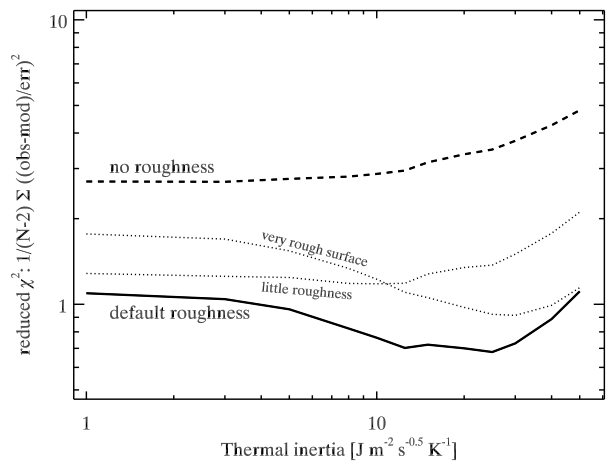}
  \caption{%
    Thermal inertia determination using the KOALA shape model and
    ground-based observations.
    We use four values of surface roughness 
    \citep[default roughness means 60\% of the surface is covered by
      craters and the RMS of the surface slopes is 0.7; see detail
      in][]{2011-PSS--Rourke}
    for thermal inertia ranging from very low to moderate
    (1--100\,J\,m$^{-2}$\,s$^{-0.5}$\,K$^{-1}$).
    \label{fig: thermal}
  }
\end{figure}

  \indent As a supplementary investigation of the capabilities of KOALA for
  the study of a large sample of asteroids from ground-based
  observations, we analyze here the thermal properties of (21)
  Lutetia, using only ground-based information:
  the KOALA 3-D shape model \citep{2010-AA-523-Drummond,
    2010-AA-523-Carry}, and  
  104 individual thermal measurements \citep[][and reference
    therein]{2011-PSS--Rourke},
  disregarding any information provided by the flyby of the asteroid
  by Rosetta.
  We then compare our results with 
  those derived from the thermal infrared observations acquired by
  the MIRO instrument during Rosetta flyby
  \citep[][]{2011-PSS--Gulkis}.
  We also comment on the results obtained by
  \citet[][]{2011-PSS--Rourke} with the same data set, but
  using the 3-D 
  shape model derived from Rosetta imaging
  \citep{2011-Science-334-Sierks}.\\
  \indent Many observations at thermal wavelengths are available:
  Lutetia was observed
  multiple times by several infrared survey missions, like
  IRAS in 1983, or
  Akari in 2006/2007,
%  or WISE in 2010/2011,
  but it was also targeted by
  different observing campaigns from
  ground: IRTF, ESO-TIMMI2,
  and from space: Spitzer, Herschel
  \citep[see][, for
    details on the observing circumstances]{2011-PSS--Rourke}.\\
  \indent First, we use the spin-vector coordinates provided by KOALA
  to determine the corresponding
  (lightcurve averaged) cross-sections for all
  observations that were used to derive 
  the absolute magnitude of Lutetia: 
  H\,=\,7.25\,$\pm$\,0.01\,mag
  \citep{1989-AsteroidsII-Bowell, 2010-AA-515-Belskaya}.
  All observations were taken with 
  sub-observer latitude between $-70$\degr~and $-85$\degr,
  close to pole-on geometry, when the  
  apparent average diameter of
  Lutetia was around 110--111\,km
  \add{\citep[compared to its volume-equivalent diameter of
      98\,km,][]{2011-Science-334-Sierks}. 
    The absolute magnitude of Lutetia was therefore slightly
    overestimated.}
  \rem{This then allows determination of}
  \add{We can still use it to determine}
  the geometric visual albedo $p_V$ via
  the relation:
  $\log p_V~=~6.2559~-~2\log d~-~0.4$H
  \citep[][and references therein]{2007-Icarus-190-Pravec},
  \rem{with $d$ the volume-equivalent spherical diameter of Lutetia}
  \add{providing $d$ is the average apparent diameter of Lutetia at
    the time of the observations}. 
  The combination of the KOALA shape model with the published
  absolute H magnitude
  leads to $p_V$\,=\,0.19\,$\pm$\,0.01,
  \add{in excellent agreement with the value 
    derived during Rosetta flyby
    \citep[0.19\,$\pm$\,0.01, see][]{2011-Science-334-Sierks}}.\\
%  in excellent agreement with the value estimated with the shape model
%  derived during Rosetta flyby
%  \citep[0.185\,$\pm$\,0.001, see][same issue]{2011-PSS--Rourke}.\\
%
  \indent Second, the knowledge of the 3-D shape
  allows \add{us to correct this bias on the actual H magnitude of
    Lutetia}:
  \rem{assignment of an ``unbiased H magnitude":}
  Published absolute magnitudes are typically derived from a limited
  number of latitudes of the sub-observer point\add{, and represent
    therefore an approximation only to the \textsl{real} absolute
    magnitude}. 
  Using the knowledge of the 3-D shape model with
  an absolute size scale and the geometric albedo (see above),
  one can determine \add{the proper} geometry-independent H-mag.
  \rem{Such an H-mag can then be considered a general,
  object-related property rather than an observed quantity valid only
  for certain geometries.}
  \add{Indeed, the absolute magnitude is intended to be a general,
  object-related property, rather than an observed quantity valid only
  for certain geometries \citep{1989-AsteroidsII-Bowell}.}
  The \rem{geometry-independent} H-mag for Lutetia, based
  on the KOALA shape model
  and the geometric visual albedo of $p_V$\,=\,0.19,
  is \add{therefore} H$_V$\,=\,7.42\,$\pm$\,0.03.\\
%
%
%
%
%
%%%%%%%%%%%%%%%%%%%%%%%%%%%%%%%%%%%%%%%%%%%%%%%%%%%%%%%%%%%%%%%%%%%%%%%%%%%%%%%%%%
%%%%%%%%%%%%%%%%%%%%%%%%%%%%%%%%%%%%%%%%%%%%%%%%%%%%%%%%%%%%%%%%%%%%%%%%%%%%%%%%%%
\begin{table}
\begin{tabular}{l@{ $\pm$ }lrl}
 \hline
 \multicolumn{2}{c}{$p_V$} & $\Gamma$ (SI) & References \\
 \hline
 \multicolumn{2}{c}{\textsl{assumed}} &   20  & \citet{2011-PSS--Gulkis} \\
 0.185 & 0.005          &    5  & \citet{2011-PSS--Rourke} \\
 0.19  & 0.01           & $<40$ & This work\\
 \hline
\end{tabular}
\caption{%
  Albedo ($p_V$), and thermal inertia ($\Gamma$, in SI units: J\,m$^{-2}$\,s$^{-0.5}$\,K$^{-1}$)
  derived using MIRO
  on-board Rosetta \citep{2011-PSS--Gulkis}, ground-based data in
  combination with the shape model derived from Rosetta flyby
  \citep{2011-PSS--Rourke}, and ground-based data in
  combination with the KOALA shape model (present study).
  \label{tab: thermal}
}
\end{table}
%%%%%%%%%%%%%%%%%%%%%%%%%%%%%%%%%%%%%%%%%%%%%%%%%%%%%%%%%%%%%%%%%%%%%%%%%%%%%%%%%%
%%%%%%%%%%%%%%%%%%%%%%%%%%%%%%%%%%%%%%%%%%%%%%%%%%%%%%%%%%%%%%%%%%%%%%%%%%%%%%%%%%
%
%
%
%
%
%
%
  \indent We use this refined H-mag to proceed with
  radiometric analysis via a TPM code
  \citep[][see
    Section~\ref{ssec: thermal}]{1996-AA-310-Lagerros,1997-AA-325-Lagerros}. 
  This model considers a 1-D heat conduction into the surface, based on
  realistic surface conditions of illumination provided by the KOALA
  shape model
  \citep[see][for details on such
    computations]{2011-PSS--Rourke}.
  As a complementary check of the techniques, we first
  determine the radiometric
  size and geometric albedo of Lutetia, regardless of the size 
  information provided by the KOALA shape model
  \citep[following the method by][]{2011-AA-525-Muller}.
  The calculation is based on the general
  thermal properties derived for large main-belt asteroids
  \citep{2002-AA-381-Muller},
  a wavelength-dependent emissivity model,
  a default thermal inertia $\Gamma$ of 15\,J\,m$^{-2}$\,s$^{-0.5}$\,K$^{-1}$,
  and a default roughness implementation
  with 60\% of the surface covered by craters
  and an RMS of the surface slopes of 0.7
  \citep[see, \eg,][for
    definitions of these quantities]{1996-AA-310-Lagerros,1997-AA-325-Lagerros}.
  This leads to a radiometric volume-equivalent diameter
  (based on a dimensionless version of the KOALA 3-D shape model) of
  99.8\,$\pm$\,4.6\,km and a geometric albedo $p_V$ of 0.198\,$\pm$\,0.017
  (weighted mean values and standard deviations from the analysis of
  the 104 individual thermal measurements).
  These values are in agreement with the pre-flyby estimates
  \add{(diameter of 98.3\,$\pm$\,5.9\,km and geometric albedo $p_V$ of
    0.208\,$\pm$\,0.025)}
  by \citet{2006-AA-447-Mueller}, using the lightcurve inversion model by
  \citet{2003-Icarus-164-Torppa}.
  This demonstrates the value of combining thermal data with
  spin and 3-D shape models to derive accurately the
  diameters and albedos of
  asteroids.\\
  \indent Finally, we constrain the thermal
  inertia of Lutetia 
  by comparing the TPM predictions
  (using the KOALA 3-D shape model, including its size estimate)
  for a range of thermal inertias with the 104 observed
  thermal measurements.
  This allows us to find the most probable
  thermal inertia to explain all data simultaneously,
  \ie, to match
  the before/after opposition observations as well as the observations
  taken at the short-wavelength Wien part of the spectral energy
  distribution (SED) and the long-wavelength Rayleigh-Jeans part of the
  SED. Thermal data also cover the entire rotation of
  Lutetia and
  a significant range of sub-Earth latitudes
  \citep[see the listing of observations in][]{2011-PSS--Rourke}.
  Thermal inertias in the
  range 5--40\,J\,m$^{-2}$\,s$^{-0.5}$\,K$^{-1}$
  produce the lowest $\chi^2$-values
  (see Fig.~\ref{fig: thermal}),
  indicating the presence of fine-grained regolith
  \citep{1987-Icarus-72-Brown}.
  These values are \rem{close to the measured values by Rosetta}
  \add{in agreement with the values measured by Rosetta} 
  \citep[$<20$\,J\,m$^{-2}$\,s$^{-0.5}$\,K$^{-1}$, see][]{2011-PSS--Gulkis}.
  The $\chi^2$-analysis also shows that one has to add 
  substantial roughness on the surface to explain the thermal
  measurements. The KOALA 3-D shape model without any small-scale
  roughness cannot provide acceptable $\chi^2$-values.\\
  \indent The prediction made using the combination of the KOALA and
  TPM models provides a good fit to all 104 measurements of Lutetia
  in the thermal infrared.
  There is a discrepancy, however, for a given observing geometry that
  could be solved by the addition of a plateau/hill to the 3-D shape
  model, whose size remains within the quoted 3-D shape uncertainty
  as discussed by \citet{2011-PSS--Rourke}.\\
  \indent From this detailed comparison of ground-based with flyby results, 
  KOALA has been shown to be a powerful technique
  for the study of asteroids from the ground, not only for size/shape/spin,
  but now also for thermal properties.
  The validated (see Section~\ref{sec: comp}) 3-D shape models,
  determined using KOALA, allow the study of the surface properties
  (\eg, albedo, thermal inertia) in great detail, and the results
  are consistent with those derived from the flyby.

%%%%%%%%%%%%%%%%%%%%%%%%%%%%%%%%%%%%%%%%%%%%%%%%%%%%%%%%%%%%%%%%%%%%%%%%%%%%%%%%%%
%%%%%%%%%%%%%%%%%%%%%%%%%%%%%%%%%%%%%%%%%%%%%%%%%%%%%%%%%%%%%%%%%%%%%%%%%%%%%%%%%%
%%%%%%%%%%%%%%%%%%%%%%%%%%%%%%%%%%%%%%%%%%%%%%%%%%%%%%%%%%%%%%%%%%%%%%%%%%%%%%%%%%
%%%%%%%%%%%%%%%%%%%%%%%%%%%%%%%%%%%%%%%%%%%%%%%%%%%%%%%%%%%%%%%%%%%%%%%%%%%%%%%%%%

\section{Concluding remarks: the future of KOALA\label{sec: conclu}}
  \indent The flyby of (21) Lutetia by ESA Rosetta provided a
  spectacular demonstration of the capabilities of the KOALA
  algorithm.
  Spin-vector coordinates were found
  to be accurate within two
  degrees, 3-D shape modeling to better than 2\,km~(local topography), 
  and dimensions to 2\%.
  Volume estimates provided by KOALA are seen to be
  accurate to \add{better than} 10\%.
  The thermal properties 
  (albedo and thermal inertia)
  of Lutetia determined using a 
  thermophysical model in conjunction with the KOALA shape model agree
  with the Rosetta-flyby-derived values, within the quoted uncertainties.
  \add{These levels of accuracy on the spin and 3-D shape/size are
    typical for large main-belt asteroids, and not specific to Lutetia.
    Although it was extensively observed from the ground,
    being a spacecraft target, the number of lightcurves was 
    similar to that of other large main-belt asteroids and 
    the geometry of imaging observations
    was not particularly favorable (\ie, mostly
    close to ``pole-on'').} \\
  \indent This ability of KOALA to determine the volume
  of \add{main-belt asteroids of size $\sim$100\,km}
  with an accuracy of about 10\%
  opens the possibility for study of a larger
  set of small bodies using Earth-based
  observations. For instance, it can be
  expected to help efforts to better understand the densities
  of asteroids belonging to
  different taxonomic classes \citep{2009-Icarus-202-DeMeo}.
  This will be assisted by mass estimates
  from about 150--200 asteroids that will be
  determined from gravitational deflections observed
  by the upcoming ESA Gaia astrometry mission \citep{2007-AA-472-Mouret},
  and the ever-growing number of \add{known} binary asteroids.
  During the first stages of the encounter
  of Rosetta with its main target, comet 67P/Churyumov-Gerasimenko
  in 2014, we also plan to use \add{a} KOALA-style
  analysis of the first, low-resolution, resolved images
  to quickly produce a shape/size model. This will support the mission
  science until a full high-resolution model can be derived.\\
  \indent \add{The} present implementation of KOALA
  allows the combined use of
  optical lightcurves (including sparse photometry),
  profiles from disk-resolved images,
  and chords from stellar occultations. 
  We continue the development of KOALA toward the
  use of more data modes
  (\eg, interferometry, thermal radiometry)
  to increase the
  number of possible targets, and \add{to} set better constraints on targets
  observable only with certain techniques.
  Using different wavelength ranges, in particular in
  the thermal infrared, opens the \add{possibility} of
  deriving additional physical
  properties, like albedo or thermal inertia.
  We also plan to incorporate additional cross-checks and constraints
  on the inversion, such as rigorous attention \add{to 
    differences in the spatial resolution of input images}, to 
  improve the confidence on the non-convex
  features and details.\\

\section*{Acknowledgments\label{sec: conclu}}

%--osiris stuff
  \indent We would like to thank the OSIRIS Team
  for use of the OSIRIS images.
  OSIRIS was built by a consortium of the Max-Planck-Institut f\"ur
  Sonnensystemforschung, Lindau, Germany, the Laboratoire
  d'Astrophysique de Marseille, France, the Centro Interdipartimentale 
  Studi e Attivita' Spaziali, University of Padova, Italy, the
  Instituto de Astrofisica de Andalucia, Granada, Spain, the Research
  and Scientific Support Department of the European Space Agency
  (ESA/ESTEC), Noordwijk, The Netherlands, the Instituto Nacional de
  Tecnica Aerospacial, Madrid, Spain, the Institut  f\"ur Datentechnik
  und Kommunikationsnetze der Technischen Universitat, Braunschweig
  and the Department of Astronomy and Space Physics of Uppsala
  University, Sweden. 
  The support of the national funding agencies DLR, CNES, ASI, MEC,
  NASA, and SNSB is gratefully acknowledged. 
  We thank the Rosetta Science Operations Center and the Rosetta
  Mission Operations Center for the successful flyby of (21) Lutetia. 
  \add{We would like to thank F. Preusker, S. Marchi,
    and J.-B. Vincent for
    providing their results ahead of publication.}
%--thermal
  Herschel is an ESA space observatory with science instruments
  provided by European-led Principal Investigator consortia and with
  important participation from NASA.
  The thermal analysis is also based on observations collected at the
  European Southern Observatory, Chile: 
  \href{http://archive.eso.org/wdb/wdb/eso/eso_archive_main/query?prog_id=79.C-0006\&max_rows_returned=1000}{79.C-0006}.
  The KOALA shape model discussed here was based on imaging observations
  realized at the ESO Very Large Telescope
  (\href{http://archive.eso.org/wdb/wdb/eso/eso_archive_main/query?prog_id=079.C-0493\%28A\%29\&max_rows_returned=1000}{079.C-0493}),
  and the W. M. Keck Observatory, which is operated as a scientific
  partnership among the California Institute of Technology, the
  University of California, and the National Aeronautics and Space
  Administration. The Observatory
  was made possible by the generous financial support of the W. M. Keck
  Foundation.
  We also thank our collaborators on Team Keck, the Keck science
  staff,
  for making possible
  some of these observations, and for observing
  time granted at Gemini Observatory under NOAO time
  allocation.
%--funds
  This work was supported, in part,
  by the NASA Planetary Astronomy and NSF
  Planetary Astronomy Programs (Merline PI), 
  and the work of J. \v{D}urech was supported by the grant GACR 209/10/0537 of the
  Czech Science Foundation.
  This research used Miriade VO tool
  \add{\citep{2008-ACM-Berthier}} developed at IMCCE, and
  NASA's Astrophysics Data System. Thanks to all the developers!




\bibliographystyle{model2-names}

%\bibliography{biblio}

\end{document}